\documentclass[twocolumn]{aa} 

\usepackage{graphicx}
\usepackage{txfonts}
\usepackage{caption}
\usepackage{subcaption}
\begin{document}

   \title{Polarimetry of the potential binary supermassive black hole system in J1430+2303\thanks{Based on observations made with ESO Very Large Telescope at the Paranal Observatory under programme ID~109.24E7.001}}
   
   \titlerunning{Polarization of J1430+2303}

   \author{F. Marin\inst{1}          
          \and
          D. Hutsem\'ekers\inst{2} 
          \and
          I. Liodakis\inst{3}
          \and          
          R. Antonucci\inst{4}           
          \and
          N. Mandarakas\inst{5,6}    
          \and
          E. Lindfors\inst{3}   
          \and
          D. Blinov\inst{5,6}             
          \and
          T. Barnouin\inst{1}   
          \and
          D. Savi\'c\inst{2}     
          }

   \institute{Universit\'e de Strasbourg, CNRS, Observatoire Astronomique de Strasbourg, UMR 7550, 11 rue de l'universit\'e, 67000 Strasbourg, France\\
             \email{frederic.marin@astro.unistra.fr}
             \and            
                 Institut d’Astrophysique et de G\'eophysique, Universit\'e de Li\`ege, All\'ee du 6 Ao\^ut 19c, B5c, 4000 Li\`ege, Belgium   
                 \and
             Finnish Centre for Astronomy with ESO, University of Turku, Quantum, Vesilinnantie 5, FI-20014, Finland              
             \and
                 University of California, Physics Department, Santa Barbara, Broida Hall, Santa Barbara, CA 93106-9530, USA 
                 \and    
                 Institute of Astrophysics, Foundation for Research and Technology-Hellas, Voutes, 71110 Heraklion, Greece      
             \and            
             Department of Physics, University of Crete, University Campus, 70013, Heraklion, Greece
             }

   \date{February 2 Day, 2023; accepted March 3, 2023}
	
 
  \abstract
   {The growth of supermassive black holes (SMBHs) through merging has long been predicted but its detection remains elusive. However, a promising target has been discovered in the Seyfert-1 galaxy J1430+2303, where two SMBHs may be about to merge.}
   {If a binary system truly lies at the center of J1430+2303, the usual symmetry expected from pole-on views in active galactic nuclei (AGNs) responsible for the observed low ($\le$ 1\%) optical linear polarization in the continuum of these objects is expected to be broken. This should lead to higher-than-usual polarization degrees, together with time-dependent variations of the polarization signal.}
   {We used the specialized photopolarimeters RoboPol mounted on the 1.3m telescope at the Skinakas Observatory and the Alhambra Faint Object Spectrograph and Camera (ALFOSC) mounted on the 2.56m Nordic Optical Telescope (NOT) at the "Roque de los Muchachos" Observatory to measure the B-, V-, R-, and I-band polarization of J1430+2303. Observations were complemented using the FORS2 spectropolarimeter mounted on the VLT to acquire 3500 -- 8650~\AA~ polarized spectra. We compared our set of observations to Monte Carlo radiative-transfer predictions to look for the presence of a SMBH binary.}
   {The observed linear continuum polarization of J1430+2303 in the V and R bands is $\sim$ 0.4\% with an associated polarization angle of slightly larger than 0$^\circ$. We detected no significant changes in polarization or photometry between May, June, and July of 2022. In addition, there is no significant difference between the polarization of H$\alpha$ and the polarization of the continuum. A single SMBH at the center of an AGN model is able to reproduce the observed spectrum and polarization, while the binary hypothesis is rejected with a probability of $\sim$85\%.}
   {The low degree of continuum polarization, the lack of variability in photometry and polarization over three months, and the absence of H$\alpha$ polarization different than that of the continuum tend to indicate that J1430+2303 is a standard Seyfert-1 AGN whose nuclear inclination is 24 -- 31$^\circ$ according to our model.}


   \maketitle
%

\section{Introduction}
\label{Introduction}

Supermassive black holes (SMBHs) are expected to grow by accretion and/or merger. Accretion is a slow process that can explain the powerful thermal emission detected in active galactic nuclei (AGNs) but detecting the presence of slowly accreting SMBHs with masses superior to 10$^9$~M$_\odot$ at redshifts $\ge$~6 \citep{Mortlock2011,Banados2018,Wang2021} is a difficult task. Although short (and therefore hardly observable) episodes of super-Eddington accretion can lead to such masses in a sufficiently short time (see, e.g., \citealt{Madau2014,Volonteri2015}), BH--BH coalescence is more likely to explain the rapid mass gain and the fast growth of most of the high-redshift SMBHs. Nevertheless, this process has never been observed. In fact, both mechanisms are likely happening throughout cosmic time, depending on the BH seed mass, environment, and redshift \citep{Pacucci2020}. Although accretion is nowadays regularly observed in AGNs, mergers of SMBHs are yet to be detected \citep{Shannon2015}. In fact, neither standard photon-based astronomy nor nanohertz gravitational wave astronomy has successfully witnessed the merger of a SMBH binary.

The answer to this fundamental question may lie in SDSS~J143016.05+230344.4 (hereafter J1430+2303). This object is a radio-quiet AGN with equatorial (J2000.0) coordinates 217.566899$^\circ$, 23.062348$^\circ$ situated in the nearby Universe ($z$ = 0.08105) and exhibits the typical emission line spectrum of a type-1 Seyfert galaxy. Recent very long baseline interferometry (VLBI) imaging confirmed the radio-quiet nature of this source, with no signs of radio outbursts or detectable jets \citep{An2022}. However, broad H$\alpha$ line from  J1430+2303  is both asymmetric and blueshifted by 2390 $\pm$ 174 km.s$^{-1}$ with respect to the narrow lines, possibly indicating a highly accreting source, in contrast to its Seyfert classification \citep{Zamfir2010}. Blueshifted emission is common for highly ionized lines, but this is certainly not the case for low-ionization lines such as H$\alpha$ for which a blueshift appears in less than 3\% of Seyfert-1s \citep{Strateva2004}. Even more intriguingly, the stochastic flux variability seen for J1430+2303, which has been known for decades, began to evolve in 2018 to enter a periodic mode, while its period and amplitude both show a uniformly decaying trend \citep{Jiang2022}. This very peculiar oscillation pattern shows up in near-infrared, optical, ultraviolet, and X-ray bands with a surprising similarity between the wavebands. The rapid decay of the period (from a year to a month within only three years) favors a scenario in which a secondary black hole orbits around the primary SMBH in an inclined, highly eccentric trajectory (\citealt{Jiang2022}, but see \citealt{Dotti2023} for an alternative scenario). According to post-Newtonian modeling of the orbital evolution, the merger is likely to occur before 2025. If true, such an event will be a first in astrophysics. This means that it is essential to observe J1430+2303 as swiftly as possible with the largest variety of techniques in order to collect information before the pre-merging phase ends.  

In this regard, polarization monitoring can be seen as a powerful tool for probing or even detecting SMBH binarity. Because polarization is enhanced by asymmetry, a system in which two BHs are orbiting around each other ---often in a noncircular orbit due to the mass difference between the two objects--- is expected to show both variable and strengthened polarization with respect to a system with only one BH \citep{Savic2019,Dotti2022}. In particular, a binary SMBH is expected to create 
periodic modulations of the linear polarization of the continuum, both in polarization degree and angle, with the minimum of the polarization degree modulation coinciding with the peak of the system's light curve \citep{Dotti2022}. If emission lines are accounted for, such as the almost omnipresent Balmer lines in the optical spectrum of face-on AGNs, complex polarization angle profiles are expected throughout the broadened lines, strongly affecting the polarized and unpolarized line profiles \citep{Savic2019}. 

For this reason, we performed polarimetric observations of J1430+2303 in the spring and summer of 2022. We present our observations in Sect.~\ref{Observation} and compare our findings to state-of-the-art modeling of AGN polarization in Sect.~\ref{Modeling}. We discuss the implication of our observations in Sect.~\ref{Discussion} before presenting our conclusions in Sect.~\ref{Conclusion}.

\section{Observations}
\label{Observation}

\begin{figure}[t]
\centering
\includegraphics[trim={1cm 0cm 0cm 1cm},clip, width=11cm]{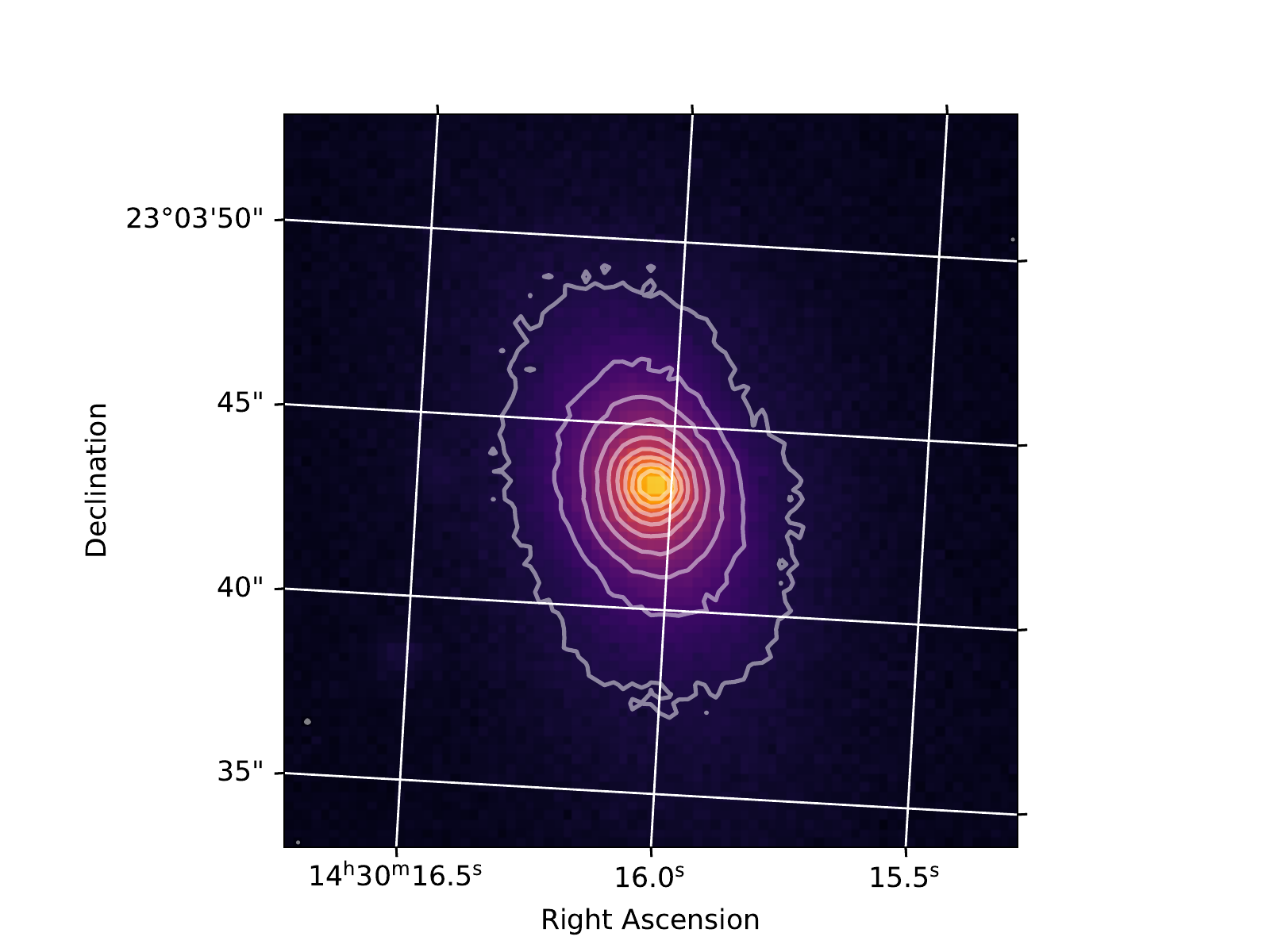}
\caption{FORS2 $R$-band acquisition image of J1430+2303. The galaxy major axis position angle is around 20$^{\rm o}$. Logarithmic isocontours are indicated in white.}
\label{fig:ima}
\end{figure}

In order to secure polarization measurements before any potential merger, we used specialized photo- and spectro-polarimeters to acquire fast and reliable broadband polarimetric measurements. For the remainder of this paper, $q$ and $u$ are the normalized Stokes parameters, $p$ is the linear polarization degree, and $\theta$ is the polarization position angle measured north to east according to the IAU convention \citep{EVPAconv}. 

J1430+2303 (Fig. \ref{fig:ima}) is located at high galactic latitude ($b$ = 66$^\circ$), in a region where the extinction is low, namely A$_V$ = 0.085 (from the NASA/IPAC Extragalactic Database\footnote{https://ned.ipac.caltech.edu/}), indicating a low interstellar polarization, $p_V \le$ 0.25\% ([$p_V$/E(B-V)]$_{\rm max}$ = 9\% mag$^{-1}$, \citealt{Serkowski1975,Panopoulou2019}). The polarization of stars at angular distances of smaller than 2$^\circ$ is low (see Table~\ref{Tab:Stars}), in agreement with the value derived from extinction. For stars at distances $d$ larger than 200~pc (which corresponds to a significant part of the Galaxy scale height in that direction), the average polarization is $p_V$ = 0.18\% $\pm$ 0.06\% with $\theta_V$ = 69$^\circ$ $\pm$ 10$^\circ$. Assuming that this value represents the interstellar polarization towards J1430+2303 at all wavelengths, all our measurements are corrected for interstellar polarization using $q_{\rm ism}$ = -0.13\% $\pm$ 0.06\% and $u_{\rm ism}$ = 0.12\% $\pm$ 0.06\%.

\begin{table}
\centering
\begin{tabular}{lccccc}
\textbf{Star} & \textbf{$\rho$ ($^\circ$)} & \textbf{$p_V$ (\%)} & {$\theta_V$ ($^\circ$)} & \textbf{$d$ (pc)} & \textbf{Ref.} \\
\hline
HD 127739    &  0.95    &   0.04 $\pm$ 0.04 &   10 $\pm$ 24 &   57  & 1 \\
BD+23 2711    &  1.40  &  0.18 $\pm$ 0.04   &  101 $\pm$ 6   &  441 & 2 \\
HD 126495    &  1.57   &  0.22 $\pm$ 0.07   &  50 $\pm$ 9   &  201 & 2 \\
BD+22 2713    &  1.60   &  0.14 $\pm$ 0.07   &  32 $\pm$ 12   &  126 & 2 \\
HD 128078    &  1.62   &  0.28 $\pm$ 0.03   &  66 $\pm$ 3   &  306 & 2 \\
\hline
\end{tabular}
\caption{Polarized stars close to J1430+2303 (references : 1 -- \citealt{Heiles2000}, 2 --  \citealt{Berdyugin2014}). $\rho$ is the angular distance to J1430+2303. Stellar distances are computed from GAIA parallaxes \citep{Jones2021}.}
\label{Tab:Stars}
\end{table}

\subsection{RoboPol}
\label{Observation:RoboPol}

We obtained optical polarimetric observations using the RoboPol instrument, which is mounted on the 1.3 m telescope at the Skinakas Observatory in Crete\footnote{https://skinakas.physics.uoc.gr/}.
RoboPol measures the linear $q$ and $u$ Stokes parameters with a single exposure by employing a combination of two fixed half-wave plates followed by two Wollaston prisms, and has no rotating parts \citep{Ramaprakash2019}. We observed J1430+2303 in three bands, that is, Johnson-Cousins R, SDSS-g, and SDSS-i, in the time period between JD~2459717.3904135 and JD~2459755.4227025. As RoboPol was originally developed to target point sources, particularly blazars \citep{Pavlidou2014}, we carefully followed a special procedure to reduce J1430+2303, it being an extended object. We measured the polarization using an elliptical aperture with its major axis aligned with the position angle of the source. The optimal size of the ellipse  varied slightly between dates and bands (presumably because of different seeing and overall environmental conditions) and was tuned to provide reliable measurements. Additionally, we masked all visible artefacts around the source to obtain as accurate a background estimation as possible. In addition to the above, the standard RoboPol analysis was performed as described in \citet{King2014}, \citet{Panopoulou2015}, and \citet{Blinov2021}. The results are shown in Table~\ref{Tab:RoboPol_Results}.

\begin{table*}
\centering
\begin{tabular}{lcccccccc}
\textbf{Date (d/m/y)} & \textbf{Aperture (")} & \textbf{Band} & \textbf{$q$ (\%)} & \textbf{$u$ (\%)} & \textbf{$p$ (\%)} & \textbf{$\sigma_{\rm p}$ (\%)} & \textbf{$\theta$ ($^{\circ}$)} & \textbf{$\sigma_{\rm \theta}$ ($^\circ$)} \\
\hline
17/05/22  &  6.8 $\times$ 5.2  &  R  &  0.86  &  0.5  &  0.99  &  0.38  &  15.1  &  9.7 \\
18/05/22  &  6.8 $\times$ 5.2  &  R  &  1.16  &  0.8  &  1.41  &  0.25  &  17.3  &  5.3 \\
18/05/22  &  5.7 $\times$ 4.4  &  sdss-i  &  1.16  &  -0.05  &  1.16  &  0.24  &  -1.2  &  5.1 \\
22/05/22  &  8.7 $\times$ 7.0  &  R  &  0.47  &  -0.25  &  0.53  &  0.36  &  -14.0  &  31.4 \\
22/05/22  &  5.7 $\times$ 4.4  &  sdss-i  &  1.27  &  0.27  &  1.30  &  0.26  &  6.0  &  4.5 \\
30/05/22  &  6.8 $\times$ 5.2  &  R  &  0.78  &  -0.58  &  0.97  &  0.18  &  -18.3  &  6.4 \\
22/06/22  &  6.8 $\times$ 5.2  &  R  &  0.18  &  -0.5  &  0.53  &  0.24  &  -41.4  &  17.1 \\
22/06/22  &  5.7 $\times$ 4.4  &  sdss-g  &  0.22  &  0.74  &  0.77  &  0.19  &  36.7  &  5.9 \\
22/06/22  &  6.8 $\times$ 5.2  &  sdss-i  &  1.64  &  0.38  &  1.68  &  0.36  &  6.5  &  5.8 \\
24/06/22  &  7.9 $\times$ 6.1  &  R  &  0.94  &  -0.46  &  1.05  &  0.33  &  -13.0  &  13.9 \\
24/06/22  &  6.8 $\times$ 5.2  &  sdss-g  &  0.63  &  0.19  &  0.66  &  0.31  &  8.4  &  18.3 \\
24/06/22  &  10.2 $\times$ 7.8  &  sdss-i  &  1.38  &  -0.92  &  1.66  &  0.66  &  -16.8  &  11.3 \\
\hline
\end{tabular}
\caption{Results from the 2022's RoboPol observations of J1430+2303.}
\label{Tab:RoboPol_Results}
\end{table*}

\subsection{NOT/ALFOSC}
\label{Observation:ALFOSC}

Observations at the Nordic Optical Telescope were performed in the $B,~V,~R,~and I$ optical bands using the Alhambra Faint Object Spectrograph and Camera (ALFOSC) on JD~2459720.61523, JD~2459724.50469, and JD~2459728.51775. The standard polarimetric setup for ALFOSC includes a half-wave plate followed by a calcite block. The instrumental polarization of the system is low $<0.3\%$. As the seeing was relatively good during the observations and we wanted to analyze the polarization of the central point source, we selected a 5'' circular aperture. The analysis was performed using the semi-automatic data-reduction pipeline of the  Tuorla Observatory. A detailed description of the pipeline and the photometric procedures used can be found in \citet{Hovatta2016} and \citet{Nilsson2018}. The estimates for the polarization degree were then debiased for the low signal-to-noise ratio following \citet{Simmons1985}. The results are shown in Table~\ref{Tab:NOT_Results} and are mostly consistent with nonsignificant detection ($p/ \sigma_p <$ 3).

\begin{table*}
\centering
\begin{tabular}{lcccccccc}
\textbf{Date (d/m/y)} & \textbf{Aperture (")} & \textbf{Band} & \textbf{$q$ (\%)} & \textbf{$u$ (\%)} & \textbf{$p$ (\%)} & \textbf{$\sigma_{\rm p}$ (\%)} & \textbf{$\theta$ ($^\circ$)} & \textbf{$\sigma_{\rm \theta}$ ($^\circ$)}  \\
\hline
20/05/22  &  5  &  B  &  0.259  &  0.281  &  0.38  &  0.51  &  23.7  &  63.3 \\
20/05/22  &  5  &  V  &  -0.138  &  0.541  &  0.56  &  0.31  &  -37.8  &  13.7 \\
20/05/22  &  5  &  R  &  -0.027  &  0.01  &  0.03  &  0.2  &  -10.2  &  63.3 \\
20/05/22  &  5  &  I  &  0.061  &  0.074  &  0.1  &  0.2  &  25.3  &  63.3 \\
24/05/22  &  5  &  B  &  0.2  &  -0.669  &  0.7  &  0.4  &  -36.7  &  23.7 \\
24/05/22  &  5  &  V  &  -0.074  &  0.601  &  0.61  &  0.23  &  -41.5  &  8.5 \\
24/05/22  &  5  &  R  &  0.046  &  -0.63  &  0.63  &  0.22  &  -42.9  &  17.7 \\
24/05/22  &  5  &  I  &  0.015  &  -0.318  &  0.32  &  0.25  &  -43.6  &  63.3 \\
28/05/22  &  5  &  B  &  0.617  &  -0.213  &  0.65  &  0.31  &  -9.5  &  21.9 \\
28/05/22  &  5  &  V  &  -0.188  &  -0.145  &  0.24  &  0.23  &  18.8  &  63.3 \\
28/05/22  &  5  &  R  &  -0.281  &  -0.014  &  0.28  &  0.17  &  1.4  &  12 \\
28/05/22  &  5  &  I  &  0.035  &  -0.398  &  0.4  &  0.19  &  -42.5  &  21.9 \\
\hline
\end{tabular}
\caption{Results from the 2022's NOT observations of J1430+2303.}
\label{Tab:NOT_Results}
\end{table*}

\subsection{VLT/FORS2}
\label{Observation:FORS2}

Spectropolarimetric observations of J1430+2303 were obtained using the European Southern Observatory (ESO) Very Large Telescope (VLT) equipped with the focal reducer/low-dispersion spectrograph FORS2 mounted at the Cassegrain focus of Unit Telescope  \#1 (Antu). Linear spectropolarimetry was performed by inserting a Wollaston prism in the beam that splits the incoming light rays into two orthogonally polarized beams separated by 22" on the CCD detector. In order to derive the normalized Stokes parameters u($\lambda$) and q($\lambda$), four frames were obtained with the half-wave plate rotated at four different position angles: 0$^\circ$, 22.5$^\circ$, 45$^\circ$, and 67.5$^\circ$. This combination allowed us to remove most of the instrumental polarization. 

Spectra were secured both with the grism 300V (blue setting, 3500-4500 \AA) alone and with the grism 300V and the order-sorting filter GG435 (red setting, 4600-8650 \AA). Observations were carried out using the blue and red settings on May 05, 2022, and only the red setting on June 29 and July 26, 2022. The slit width was 0.7" on the sky, providing us with an average resolving power of R $\approx$ 630. The slit was positioned along the parallactic angle. CCD pixels were binned 2 $\times$ 2, which corresponds to a spatial scale of 0.25" per binned pixel. At all epochs, the airmass was between 1.5 and 1.8, the seeing $\le$ 0.9", and the sky clear. Polarized (Vela1 95 = Ve 6-23, Hiltner 652) and unpolarized (HD 97689, WD 1615-154) standard stars \citep{Fossati2007} were observed on the nights of observation, or two nights before in the worst case.

\begin{figure*}
\centering
\includegraphics[width=0.7\textwidth]{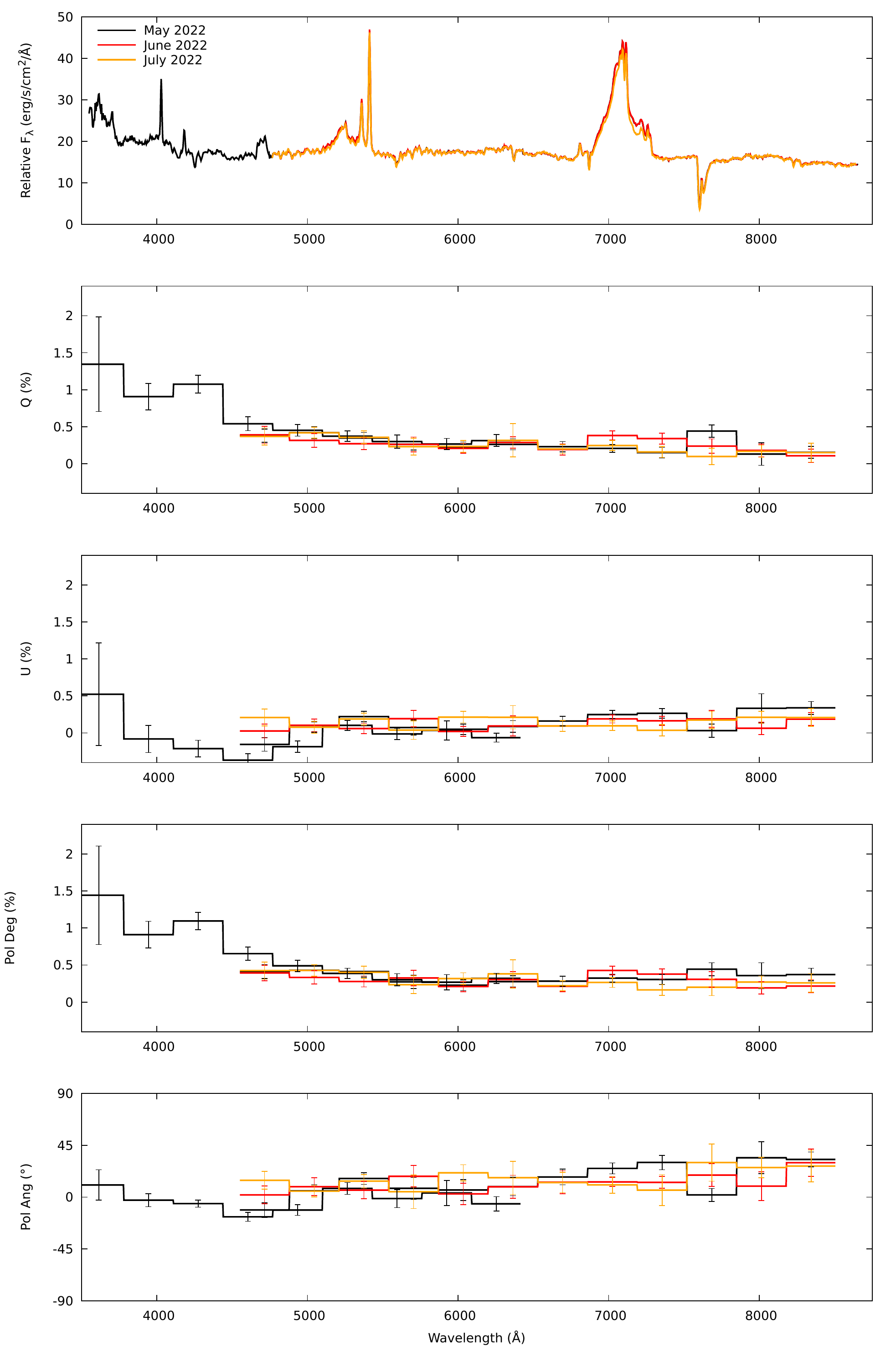}
\caption{VLT/FORS2 spectropolarimetry of J1430+2303 obtained on May 29 (black), June 29 (red), and July 24, 2022 (orange). The polarization data are rebinned using the median of 100 spectral elements.}
\label{Fig:VLTs_pectra}
\end{figure*}

Raw frames were first processed to remove cosmic-ray hits using the Python implementation of the "lacosmic" package \citep{Dokkum2001,Dokkum2012}. The ESO FORS2 pipeline \citep{Izzo2019} was then used to obtain images with two-dimensional spectra rectified and calibrated in wavelength. The one-dimensional spectra were extracted using several apertures centered on the nucleus, ranging from 0.75" to 3.5" along the spatial direction. Smaller apertures minimize the contamination by the host-galaxy stellar light but lead to spurious results in the blue setting, which is likely due to variable seeing and higher atmospheric extinction. We finally adopted an extraction aperture of 3.25" (13 pixels), which is roughly three times the seeing value. The final polarization measurements are stable with respect to small changes of the extraction aperture. The sky spectrum was estimated from adjacent MOS strips and subtracted from the nucleus spectrum. The normalized Stokes parameters u($\lambda$) and q($\lambda$) obtained with respect to the parallactic direction were subsequently rotated to the standard north--south direction. The polarization degree p($\lambda$) and polarization angle $\theta$($\lambda$) were then computed using the standard formulae. The direct spectrum F($\lambda$) was calibrated in flux using a master response curve that does not include the effect of the polarization optics. The quantities q($\lambda$), u($\lambda$), p($\lambda$), and $\theta$($\lambda$), on the other hand, are independent of the flux calibration. Resulting spectra are illustrated in Fig.~\ref{Fig:VLTs_pectra} and polarization measurements integrated over specific spectral ranges corresponding to the continuum or to spectral lines are presented in Table~\ref{Tab:VLT_Results}. The instrumental polarization estimated from the unpolarized standard stars is lower than 0.1\%. 

\begin{table*}
\centering
\begin{tabular}{lcccccccc}
\textbf{Date (d/m/y)} & \textbf{Setting} & \textbf{Spectral range (\AA)} & \textbf{$q$ (\%)} & \textbf{$u$ (\%)} & \textbf{$p$ (\%)} & \textbf{$\sigma_{\rm p}$ (\%)} & \textbf{$\theta$ ($^\circ$)} & \textbf{$\sigma_{\rm \theta}$ ($^\circ$)} \\
\hline
29/05/22        &       Blue    &       3500 - 6550 (Full)      &       0.74    &       -0.15   &       0.75    &       0.08    &       174     &       3       \\
29/05/22        &       Blue    &       3510 - 5390 (B) &       0.93    &       -0.16   &       0.95    &       0.1     &       175     &       3       \\
29/05/22        &       Blue    &       4630 - 6390 (V) &       0.46    &       -0.15   &       0.49    &       0.07    &       171     &       4       \\
29/05/22        &       Red     &       4600 - 8650 (Full)      &       0.38    &       0.02    &       0.38    &       0.06    &       2       &       5       \\
29/05/22        &       Red     &       4630 - 6390 (V) &       0.45    &       -0.04   &       0.45    &       0.07    &       177     &       4       \\
29/05/22        &       Red     &       5200 - 7960 (R) &       0.41    &       0.05    &       0.42    &       0.06    &       3       &       4       \\
29/05/22        &       Red     &       5700 - 6700 (Continuum)         &       0.42    &       0       &       0.42    &       0.07    &       180     &       5       \\
29/05/22        &       Red     &       6850 - 7300 (H$\alpha$) &       0.37    &       0.13    &       0.39    &       0.07    &       9       &       5       \\
29/05/22        &       Red     &       7020 - 7130 (H$\alpha$ Core)    &       0.24    &       0.22    &       0.33    &       0.09    &       21      &       7       \\
\hline
29/06/22        &       Red     &       4600 - 8650 (Full)      &       0.42    &       -0.02   &       0.42    &       0.06    &       178     &       4       \\
29/06/22        &       Red     &       4630 - 6390 (V) &       0.42    &       -0.05   &       0.42    &       0.07    &       176     &       5       \\
29/06/22        &       Red     &       5200 - 7960 (R) &       0.41    &       -0.01   &       0.41    &       0.06    &       180     &       4       \\
29/06/22        &       Red     &       5700 - 6700 (Continuum)         &       0.36    &       -0.03   &       0.36    &       0.07    &       178     &       6       \\
29/06/22        &       Red     &       6850 - 7300 (H$\alpha$) &       0.49    &       0.03    &       0.49    &       0.07    &       2       &       4       \\
29/06/22        &       Red     &       7020 - 7130 (H$\alpha$ Core)    &       0.54    &       -0.11   &       0.55    &       0.09    &       174     &       5       \\
\hline
26/07/22        &       Red     &       4600 - 8650 (Full)      &       0.34    &       0.04    &       0.34    &       0.06    &       3       &       5       \\
26/07/22        &       Red     &       4630 - 6390 (V) &       0.34    &       0.08    &       0.35    &       0.07    &       7       &       6       \\
26/07/22        &       Red     &       5200 - 7960 (R) &       0.3     &       0.04    &       0.31    &       0.07    &       4       &       6       \\
26/07/22        &       Red     &       5700 - 6700 (Continuum)         &       0.29    &       0.16    &       0.33    &       0.08    &       14      &       7       \\
26/07/22        &       Red     &       6850 - 7300 (H$\alpha$) &       0.36    &       -0.14   &       0.39    &       0.08    &       170     &       6       \\
26/07/22        &       Red     &       7020 - 7130 (H$\alpha$ Core)    &       0.4     &       -0.1    &       0.42    &       0.1     &       173     &       7       \\
\hline
Average &       Red     &       4600 - 8650 (Full)      &       0.38    &       0.01    &       0.38    &       0.06    &       1       &       5       \\
Average &       Red     &       4630 - 6390 (V) &       0.4     &       0       &       0.4     &       0.06    &       180     &       4       \\
Average &       Red     &       5200 - 7960 (R) &       0.38    &       0.03    &       0.38    &       0.06    &       2       &       5       \\
Average &       Red     &       5700 - 6700 (Continuum)         &       0.36    &       0.04    &       0.36    &       0.06    &       3       &       5       \\
Average &       Red     &       6850 - 7300 (H$\alpha$) &       0.41    &       0.01    &       0.41    &       0.06    &       0       &       5       \\
Average &       Red     &       7020 - 7130 (H$\alpha$ Core)    &       0.39    &       0       &       0.39    &       0.07    &       0       &       5       \\
\hline
\end{tabular}
\caption{Results from the 2022 VLT/FORS2 observations of J1430+2303 corresponding to what is shown in Fig.~\ref{Fig:VLTs_pectra}. The integrated polarization measurements correspond to a rectangular aperture of 3.25" $\times$ 0.7" (the slit width) centered on the AGN. These measurements are corrected for interstellar polarization.}
\label{Tab:VLT_Results}
\end{table*}

\subsection{Global analysis of the observed polarization}
\label{Observation:analysis}

\subsubsection{Photometric and spectral variability}
\label{Observation:analysis:flux}
\begin{figure}
\includegraphics[trim={0 0 0cm 0.25cm},clip, width=8.8cm]{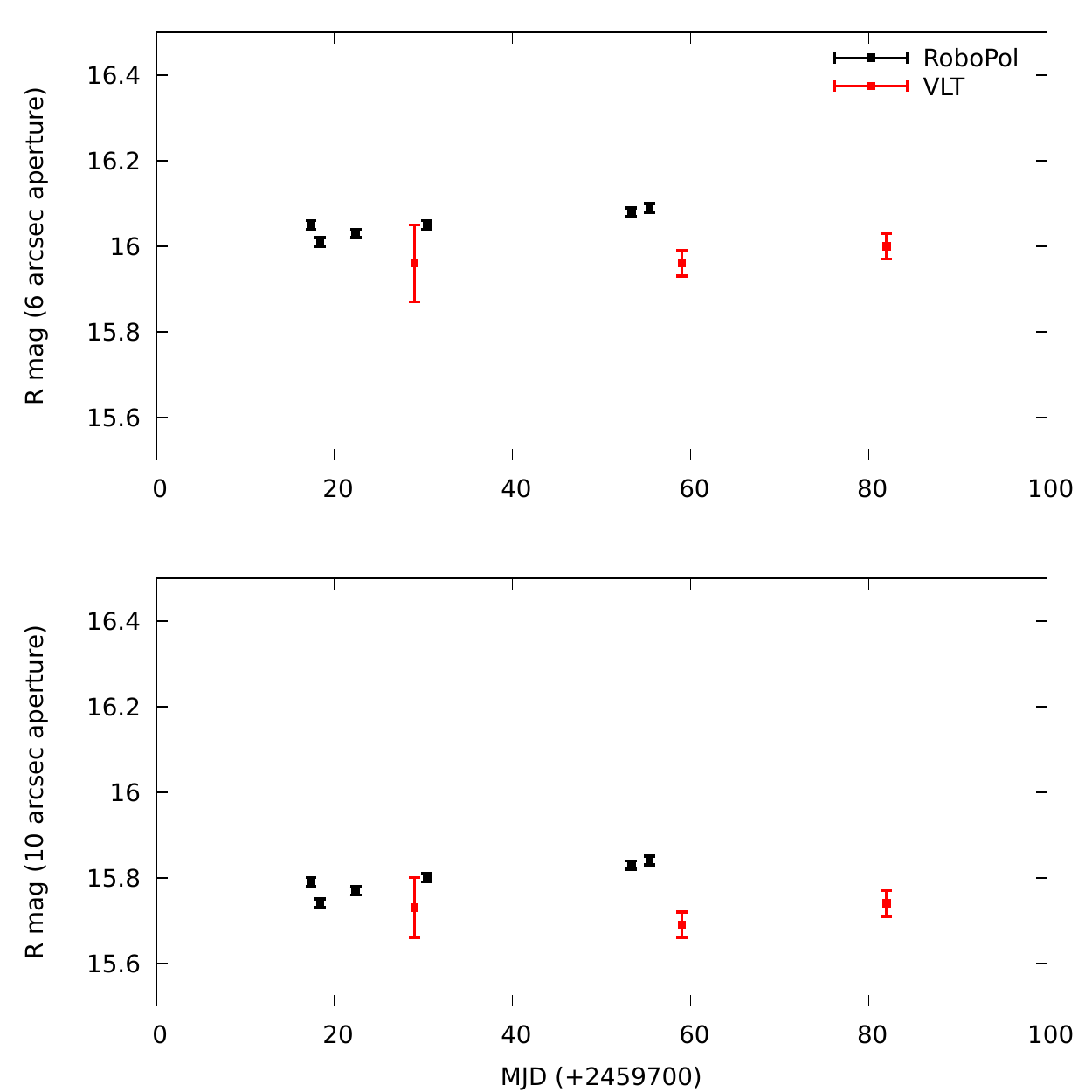}
\caption{Time-dependent R band magnitude of J1430+2303 using two different apertures (top : 6", bottom : 10").}
\label{Fig:magnitude}
\end{figure}

From the acquisition images at the VLT/FORS2 and RoboPol, we measured the R magnitude of J1430+2303 at the different epochs using two apertures (Table~\ref{Tab:Photometry}). For Robopol, we used a random field star as a reference for differential photometry, while its magnitude was transformed from sdss dr8 to the cousins-R band. During the VLT observations, the sky was clear and the photometry was performed using the zero points and extinction measurements available nightly from the FORS2 calibration plan. As it can be seen in Fig.~\ref{Fig:magnitude}, within the uncertainties of the instrument and of the reference star, the brightness of J1430+2303 does not change between the various epochs. 

Spectra obtained by the VLT/FORS2 at the three different epochs are compared in Fig.~\ref{Fig:VLTs_pectra}. The spectra obtained in June and July are normalized to the continuum of the May spectrum. The three spectra can be almost perfectly superimposed on one another, except for the H$\alpha$ broad line, which is marginally fainter in the July spectrum: the intensity of the line decreases in its central and red parts by $\sim$ 11\%, while the blue side of the line remains unperturbed. We note that subtle but significant differences can also be observed with respect to the January spectrum shown by \citet{Jiang2022}. The moderate change in the red part of the Balmer line with respect to the absence of variation in the blue part could naturally be explained by line emission from optically thin gas clouds in Keplerian motion with a non-null radial component (moving inward), as expected from our current knowledge of the broad emission-line region (BLR, see, e.g., \citealt{Gaskell2009}).

\begin{table}
\centering
\begin{tabular}{lccc}
\textbf{Date} & \textbf{Instrument} & \textbf{Aperture (")} & \textbf{R (mag)} \\
\hline
17/05/22        &       RoboPol &       6       &       16.05   $\pm$   0.01    \\
~       &       RoboPol &       10      &       15.79   $\pm$   0.01    \\
18/05/22        &       RoboPol &       6       &       16.01   $\pm$   0.01    \\
~       &       RoboPol &       10      &       15.74   $\pm$   0.01    \\
22/05/22        &       RoboPol &       6       &       16.03   $\pm$   0.01    \\
~       &       RoboPol &       10      &       15.77   $\pm$   0.01    \\
29/05/22        &       VLT     &       6       &       15.96   $\pm$   0.09    \\
~       &       VLT     &       10      &       15.73   $\pm$   0.07    \\
30/05/22        &       RoboPol &       6       &       16.05   $\pm$   0.01    \\
~       &       RoboPol &       10      &       15.8    $\pm$   0.01    \\
22/06/22        &       RoboPol &       6       &       16.08   $\pm$   0.01    \\
~       &       RoboPol &       10      &       15.83   $\pm$   0.01    \\
24/06/22        &       RoboPol &       6       &       16.09   $\pm$   0.01    \\
~       &       RoboPol &       10      &       15.84   $\pm$   0.01    \\
29/06/22        &       VLT     &       6       &       15.96   $\pm$   0.03    \\
~       &       VLT     &       10      &       15.69   $\pm$   0.03    \\
26/07/22        &       VLT     &       6       &       16      $\pm$   0.03    \\
~       &       VLT     &       10      &       15.74   $\pm$   0.03    \\
\hline
\end{tabular}
\caption{R-band photometry of J1430+2303.}
\label{Tab:Photometry}
\end{table}

\subsubsection{Continuum polarization}
\label{Observation:analysis:continuum}

From the VLT/FORS2 data, the average polarization degree is $p$ = 0.40\% $\pm$ 0.06\%, $\theta$ = 0$^\circ$ $\pm$ 4$^\circ$ in the V band, and $p$ = 0.38\% $\pm$ 0.06\%, $\theta$ = 2$^\circ$ $\pm$ 5$^\circ$ in the R band. The two are statistically the same and there are no significant changes between the three epochs. Although low, the polarization is higher than the instrumental polarization. 

\begin{figure*}
\centering
\begin{minipage}{.5\textwidth}
  \centering
  \includegraphics[width=8.8cm]{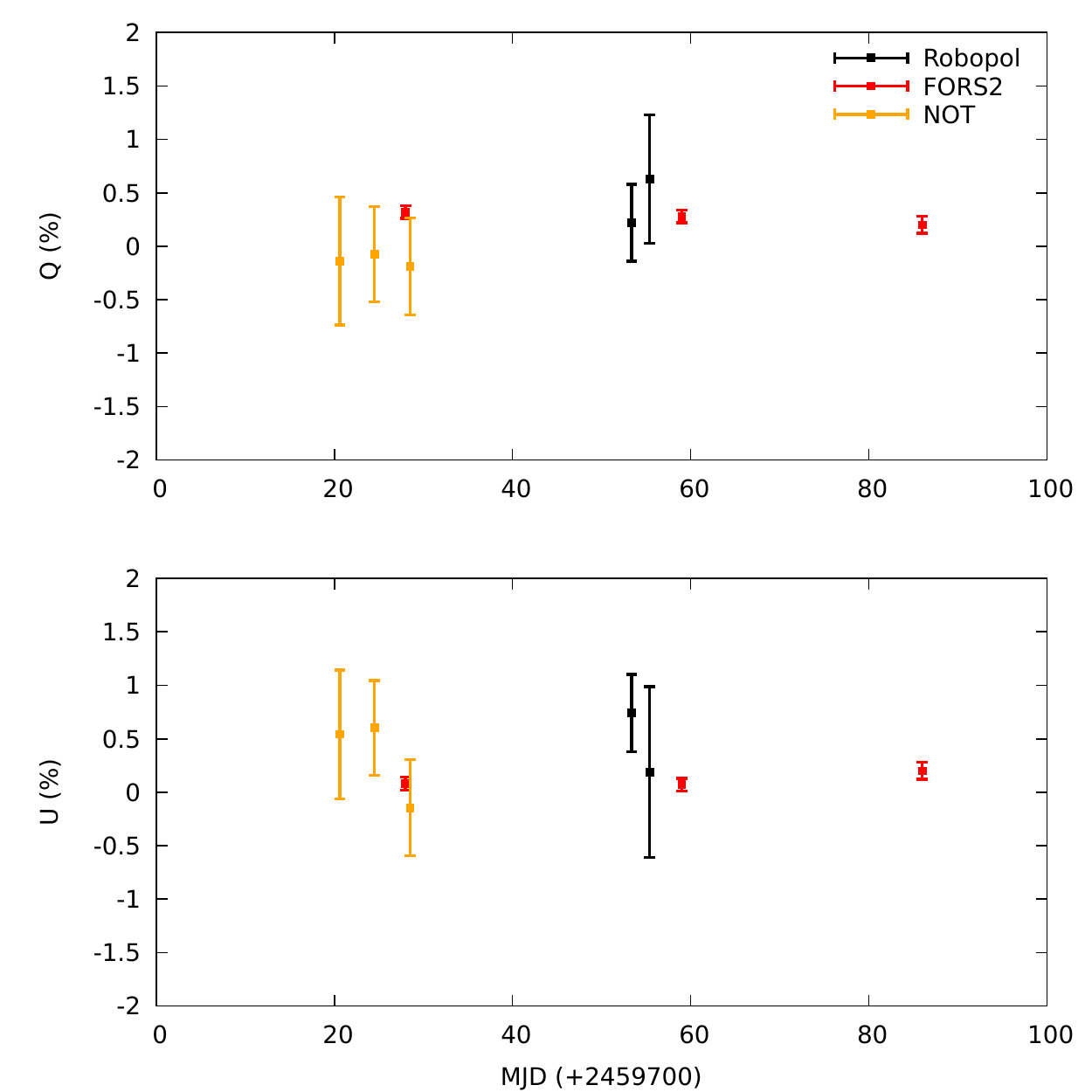}
  \label{fig:test1}
\end{minipage}%
\begin{minipage}{.5\textwidth}
  \centering
  \includegraphics[width=8.8cm]{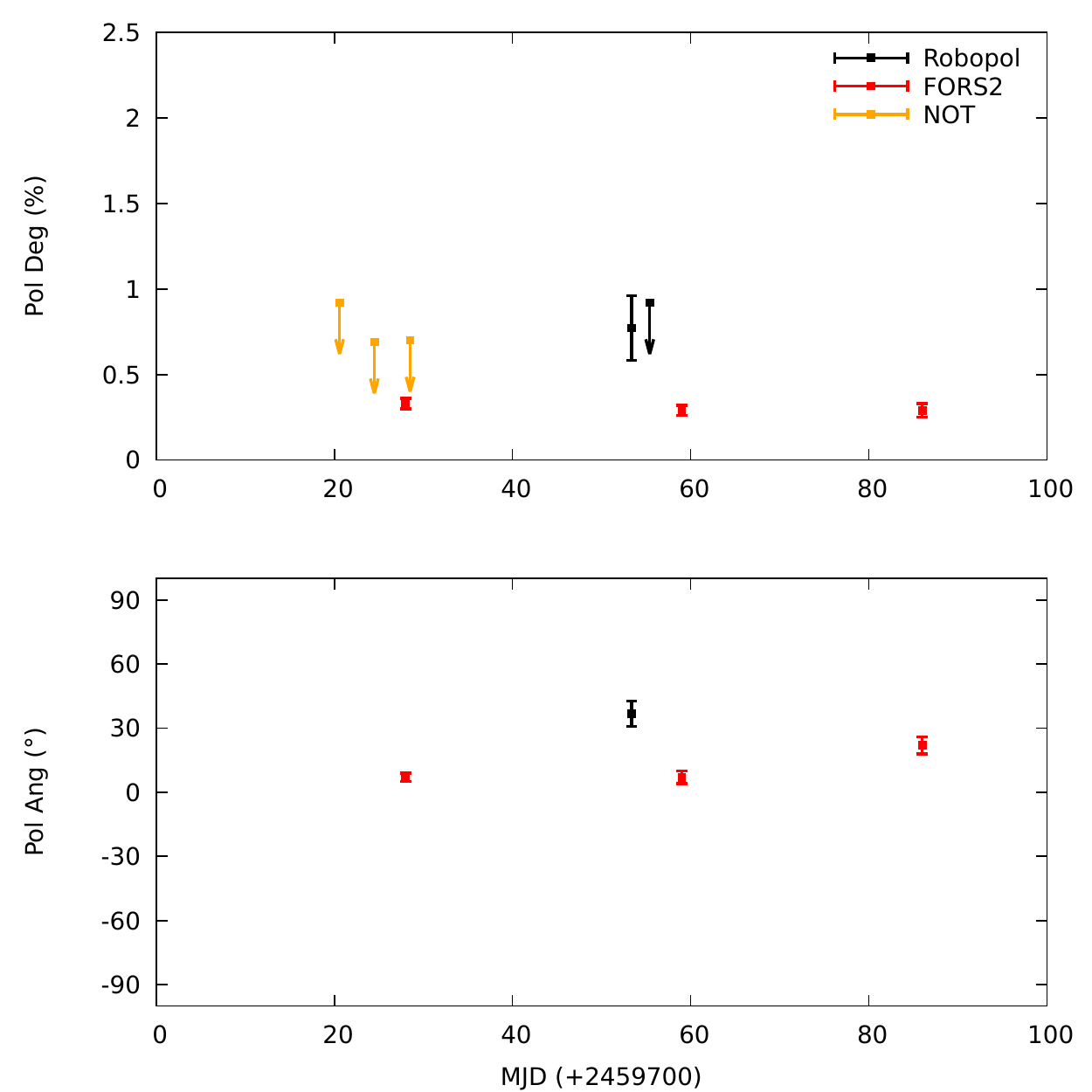}
  \label{fig:test2}
\end{minipage}
\caption{Time-dependent V band (= SDSS g) Stokes parameters (left column, top : Q, bottom : U) and polarization (right column, top: polarization degree; bottom: polarization angle) of J1430+2303. Error bars are 2 sigma and upper limits are 3 sigma.}
\label{Fig:V_pol}
\end{figure*}

\begin{figure*}
\centering
\begin{minipage}{.5\textwidth}
  \centering
  \includegraphics[width=8.8cm]{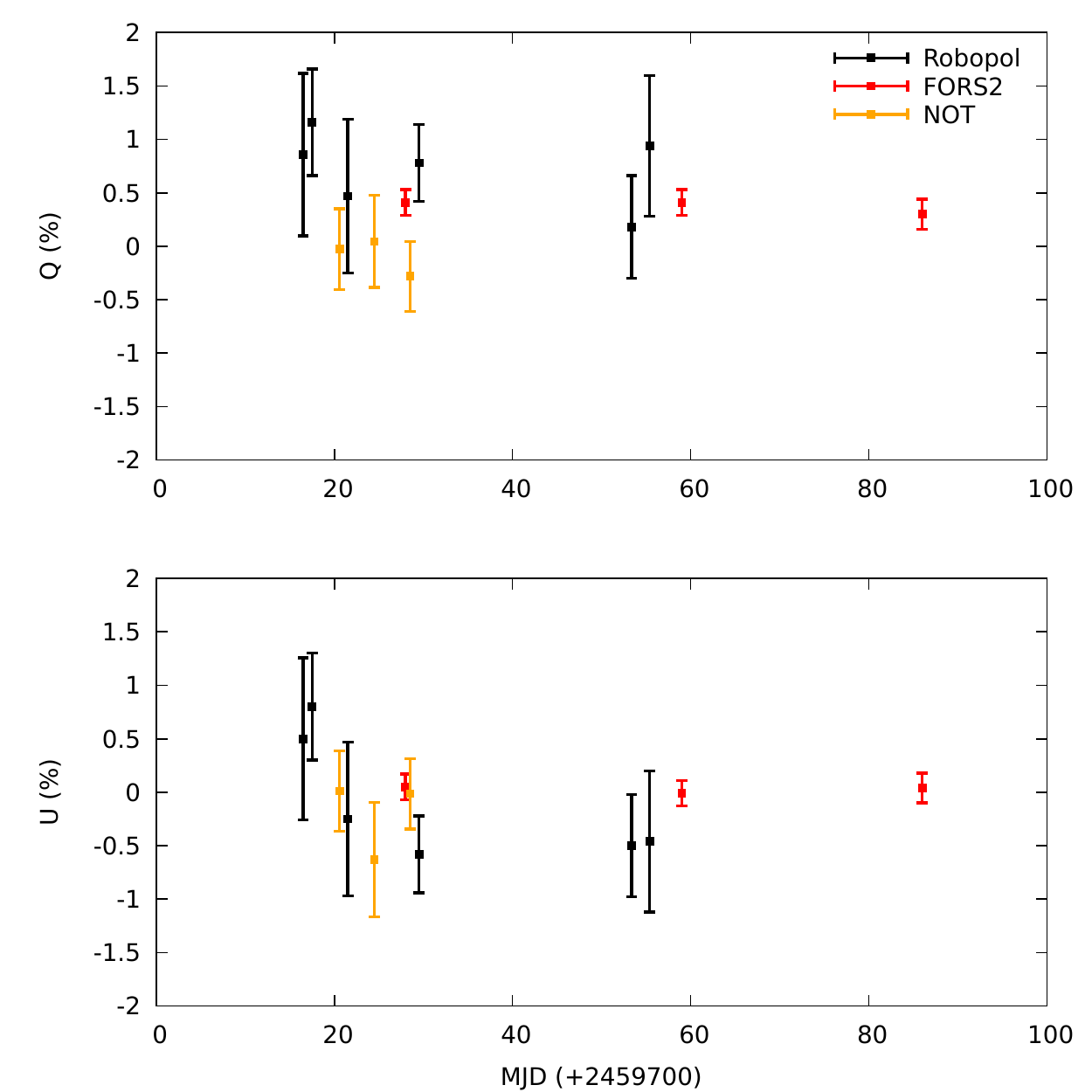}
  \label{fig:test1}
\end{minipage}%
\begin{minipage}{.5\textwidth}
  \centering
  \includegraphics[width=8.8cm]{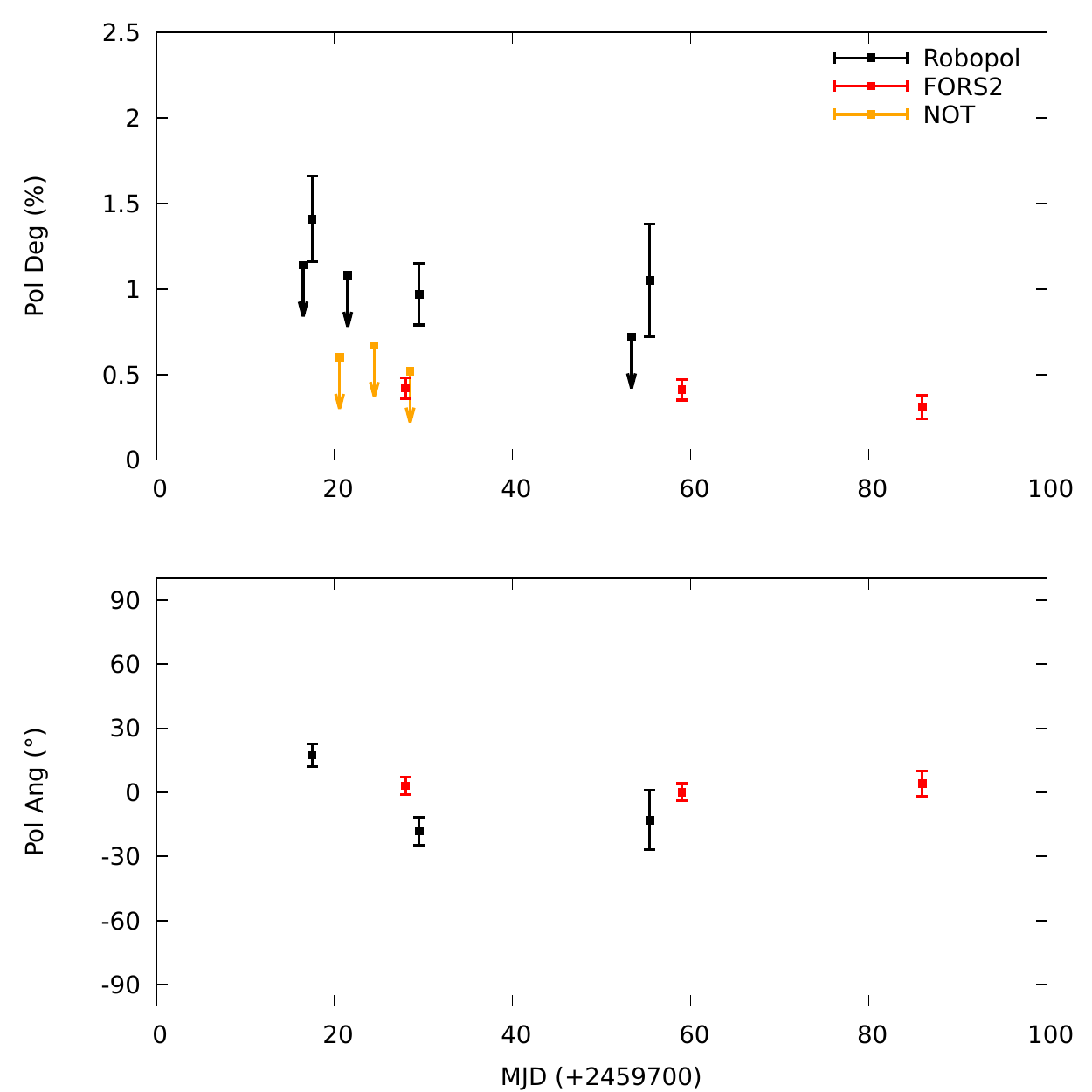}
  \label{fig:test2}
\end{minipage}
\caption{Same as Fig.~\ref{Fig:V_pol} but for the R band.}
\label{Fig:R_pol}
\end{figure*}

The VLT/FORS2 polarization measurements appear slightly different than the ones acquired by RoboPol and NOT (see Tables~\ref{Tab:RoboPol_Results}, \ref{Tab:NOT_Results}, and \ref{Tab:VLT_Results}). To check whether this can come from a difference in instrument aperture, we carried out spectrum-extraction tests by increasing the VLT/FORS2 slit length from 3.25" to 6.25". We find that the degree of polarization measured by the VLT/FORS2 decreases with increasing aperture because of the higher diluting fraction of starlight from the host galaxy, as expected from past observations of many sources \citep[e.g.,][]{Andruchow2008,Marin2018,Blinov2021}, but this does not help to explain why RoboPol and NOT polarization data seem to be slightly larger than the VLT's. We therefore examined the broad-band V/sdss-g and R data considering only the Stokes parameters $q$ and $u$, and not $p$ (see Figs.~\ref{Fig:V_pol} and \ref{Fig:R_pol}, left columns); thus, bias and upper limits are avoided and values can be directly compared with their errors (see right columns of the aforementioned figures). We find that the RoboPol and NOT measurements are almost all in agreement with the VLT measurements within 2 sigma. By calculating the average of $q$ and $u$ in RoboPol plus NOT, the final values are in agreement with those of the VLT. In conclusion, the apparent differences between VLT, RoboPol, and NOT measurement are simply noise. For this reason, we only consider VLT/FORS2 data in the remainder of the article, as they are the most precise (larger integration time, lower instrumental polarization, larger telescope).

Nevertheless, in all cases, the polarization we observed in the B band is higher than in the V and R bands, with a comparable polarization angle. The polarization angle is close to the host galaxy position angle ($\approx$ 20$^\circ$, see Fig.~\ref{fig:ima}). 

\subsubsection{Dilution by the host galaxy}
\label{Observation:analysis:host}

Assuming that the host galaxy is an old elliptical \citep{Jiang2022}, we can estimate the stellar light contribution to the spectrum of J1430+2303 at some wavelengths using the equivalent width of strong stellar features such as  CaII~H\&K ($\lambda_{\rm obs}$ = 4270~\AA) or MgI~b ($\lambda_{\rm obs}$ = 5590~\AA) measured in both the object spectrum and a galaxy template. The fraction of stellar light $f_{sl}$ in the total AGN + host light is given by $f_{sl}$ = $W_T$/$W_G$ , where $W_T$ is the equivalent width of a stellar feature in the spectrum of J1430+2303, and $W_G$ the equivalent width of the same feature measured in the galaxy template. For the galaxy template, we use the 13 Gyr Elliptical galaxy of the SWIRE Template Library \citep{Polletta2007}. The contribution of the stellar light to the V band can be estimated using the equivalent width of MgI~b. We find  $f_{sl} \approx$ 15\% at all epochs. Assuming the host-galaxy light is unpolarized, the AGN intrinsic polarization estimated using $p_{\rm agn}$ = $p_{\rm measured}$/(1 - $f_{sl}$) is only 20\% higher than the measured polarization in the V band. Even accounting for this extra source of depolarization, the observed linear polarization degree of the continuum is $\lesssim$ 1\%, as expected from standard, single-SMBH, type-1 AGNs \citep[see, e.g.,][]{Smith2002}.

The fact that the continuum polarization angle is roughly parallel to the major axis of the 
host galaxy may indicate that the host-galaxy light is slightly polarized. However, we also note that parallel polarization to the axis of the  galaxy, especially in redder bands, can be due to dichroic absorption of dust aligned perpendicularly to the presumable large-scale magnetic field of the galaxy \citep[see, e.g.,][]{Packham2011,Lopez-Rodriguez2013}. Infrared polarimetry is needed to investigate this potential mechanism.

\subsubsection{H$\alpha$ polarization}
\label{Observation:analysis:Halpha}

There is no significant difference between the polarization of H$\alpha$ and the polarization of the continuum; see Fig.~\ref{Fig:VLTs_pectra} and Table~\ref{Tab:VLT_Results}. Although shown with only two bins to maximize the signal-to-noise ratio in Fig.~\ref{Fig:VLTs_pectra}, the $q$ and $u$ spectra at natural VLT/FORS2 resolution (3~\AA, not shown here) have similar values in the line and the continuum too. It follows that any synchrotron origin for the observed polarization is ruled out and that scattering is responsible for both H$\alpha$ and continuum polarizations \citep{Angel1976}. A key point here is the lack of variability in polarization in both the continuum and the Balmer lines, which argues against a nonthermal synchrotron origin.

\section{Modeling}
\label{Modeling}
To determine whether the degree of linear polarization we observe can be produced by a system with a single and/or a binary SMBH, we ran Monte Carlo radiative-transfer simulations using the {\sc stokes} code \citep{Goosmann2007,Marin2012,Marin2015,Rojas2018,Marin2018b}. {\sc stokes} is able to simulate emission, scattering, and absorption of photons in a three-dimensional space in order to compute the outgoing polarization in the near-infrared to X-ray band. The code has been extensively used to examine the polarization of radio-quiet AGNs and we ran it for various configurations.

\subsection{A radio-quiet AGN with a single SMBH}
\label{Modeling:Single_AGN}
First, we simulate a typical radio-quiet AGN using the standard recipe already explored in \citet{Marin2012}: a central SMBH is surrounded by its geometrically thin, optically thick accretion disk that emits thermal (unpolarized) radiation. Around the core is a photo-ionized region in fast Keplerian motion that is responsible for the emission of broad emission lines, that is, the BLR. The outer radius of the BLR mixes with the optically thick dusty reservoir that blocks the view of observers situated along the equatorial plane, the so-called torus. Finally, the torus funnel collimates the disk ejection winds along the polar direction. This is the classical description of a radio-quiet AGN \citep{Antonucci1993}. The physical parameters of the model(s) are summarized in Table~\ref{Tab:AGN_model} (see \citealt{Marin2012,Marin2018b} for additional details).

\begin{table}
\centering
\begin{tabular}{lc}
\textbf{Continuum source (disk)} & ~ \\
Spectral slope $\alpha$    &   1    \\
Unpolarized continuum & ~ \\
\hline
\textbf{Accretion disk} & ~ \\
R$_{\rm in}$ & 10$^{-5}$ pc \\
R$_{\rm out}$ & 10$^{-3}$ pc \\
Width & 10$^{-4}$ pc \\
$\tau_e$ &  $>>$ 1 \\
\hline
\textbf{BLR (flared disk)} & ~ \\
R$_{\rm in}$ & 10$^{-3}$ pc \\
R$_{\rm out}$ & 10$^{-1}$ pc \\
$\theta_{\rm op}$ & 70$^\circ$ \\
$\tau_e$ & 0.1 \\
\hline
\textbf{Torus (flared disk)} & ~ \\
R$_{\rm in}$ & 10$^{-1}$ pc \\
R$_{\rm out}$ & 5 pc \\
$\theta_{\rm op}$ & 30$^\circ$ -- 60$^\circ$ \\
$\tau_d$ &  50 \\
\hline
\textbf{Polar outflows (cones)} & ~ \\
R$_{\rm in}$ & 10$^{-2}$ pc \\
R$_{\rm out}$ & 10 pc \\
$\theta_{\rm op}$ & 30$^\circ$ -- 60$^\circ$ \\
$\tau_e$ &  0.03 \\
\end{tabular}
\caption{Input parameters for the AGN model used in the {\sc stokes} simulations.
R$_{\rm in}$ and R$_{\rm out}$ are the inner and outer radii of the region, respectively.
$\theta_{\rm op}$ is the half-opening angle of the region, defined with respect to the vertical axis of the system. $\tau_e$ and $\tau_d$ are the electron and dust optical thickness, respectively. The dust model follows the standard "Milky Way" prescription \citep{Weingartner2001}.}
\label{Tab:AGN_model}
\end{table}

Several key parameters govern the polarization state of escaping photons. In particular, the half-opening angle of the torus plays an important role in determining the fraction of type-1 (pole-on) versus type-2 (edge-on) AGNs, and its value likely lies between 30$^\circ$ and 60$^\circ$ \citep{Marin2014}. We therefore use {\sc stokes} to model several AGN morphologies by varying the torus half-opening angle (30$^\circ$, 45$^\circ$ and 60$^\circ$) and look at the outgoing polarization at all possible viewing angles.

\begin{figure}
\includegraphics[trim={0 0 0.3cm 0.25cm},clip, width=8.8cm]{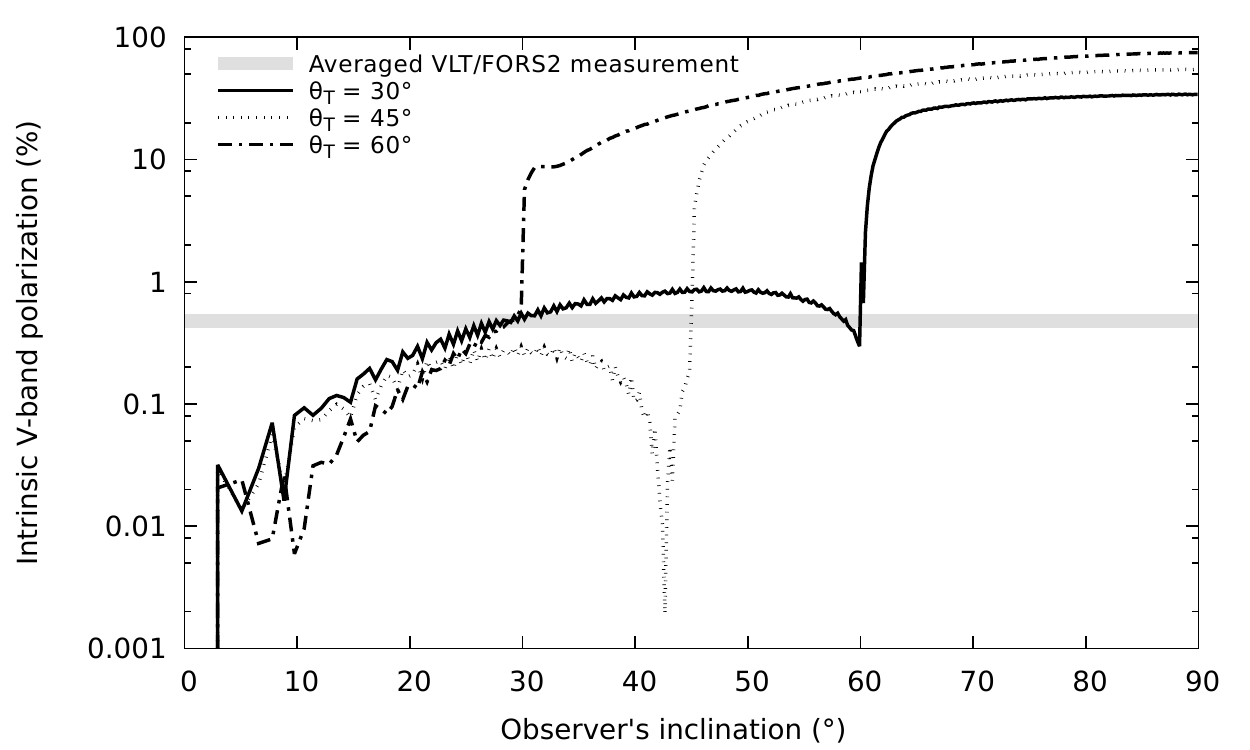}
\caption{Monte-Carlo radiative transfer modeling of the intrinsic V-band polarization 
emerging from a Seyfert AGN with only one supermassive black hole at its center.
Three models, with varying torus half-opening angles (30$^\circ$, 45$^\circ$ and 60$^\circ$, 
with respect to the vertical axis of the system) are shown. In all cases, the ionization
cones fill the solid angle funneled by the torus. The averaged V-band VLT/FORS2 polarimetric measurement, corrected for host dilution (0.48\% $\pm$ 0.06\%), is reported in gray.}
\label{Fig:Single_BH_modeling}
\end{figure}

Results are shown in Fig.~\ref{Fig:Single_BH_modeling} for the V-band continuum linear polarization of a standard AGN model. The simulated polarization originates from multiple scattering in the accretion disk, in the BLR, in the torus, and in the winds. The chaotic behavior of the polarization degree for very small inclinations is due to numerical noise, because the different viewing angles in {\sc stokes} are equally distributed in cos($\theta_i$), with $\theta_i$ being the viewing angle of the observer. As expected from past numerical modeling, the polarization degree is very low for inclinations lower than the torus horizon (type-1 view), and then rises as soon as the central engine becomes obscured (type-2 view). The small (p $\le$ 1\%) polarization detected for type-1s results from the additive dilution by the central unpolarized source and the symmetry of the system. However, when the central engine is hidden by the circumnuclear dust reservoir, photons can only escape by perpendicular scattering off the polar winds. The resulting polarization increases (as it is a function of the scattering angle), but the polarization position angle rotates by 90$^\circ$. This rotation can be seen in Fig.~\ref{Fig:Single_BH_modeling}, when $p$ is almost completely depolarized just before the type-2 regime. 

Although we do not know the orientation of the polarization position angle of J1430+2303 with respect to the reference axis (the sub-parsec jet position angle, which is currently unconstrained; see \citealt{An2022}), we can compare the modeled polarization degree to the observed intrinsic one. We can see from Fig.~\ref{Fig:Single_BH_modeling} that the VLT/FORS2 continuum linear polarization matches the three AGN models at inclinations 24$^\circ$ -- 31$^\circ$, depending on the torus--wind half-opening angle of the system. This is consistent with type-1 inclinations and, at first glance, no binary SMBHs are required to reproduce the measured continuum polarization of J1430+2303.

\subsection{Binary SMBH systems}
\label{Modeling:Binary_SMBH}
If two SMBHs lie inside the torus of J1430+2303, the symmetry of the system becomes broken and higher polarization may arise for type-1 observers. To explore this possibility, we investigate the potential parameter space of the interaction of two SMBHs. Indeed, the two compact objects (and their respective accretion disks) may not share the same orientation with respect to the torus axis, and perhaps not even the same orientation with respect to each other. We therefore model the thermal emission and reprocessing (electron scattering) within the optically thick, geometrically thin accretion disk detailed in Table~\ref{Tab:AGN_model} for all possible inclinations (0 -- 90$^\circ$). Put simply, our model assumes two SMBHs, each with its own emitting and scattering accretion disk, orbiting around each other with randomly oriented axes and inclinations with respect to the observer. Surrounding this binary component are a single BLR region, an associated torus, and polar outflows parameterized as in Table~\ref{Tab:AGN_model}. Only the outer radius of the two accretion disks has been reduced by a factor 10 in order to fit them within the BLR central "hole". Indeed, \citet{Jiang2022} estimated that the size of the circumbinary BLR in their model was $\ll$ 7 light days, allowing the two black holes to orbit each other without penetrating the BLR.

This is the simplest model for a binary system that has just begun coalescence, before the alignment of the structures associated with the two SMBHs \citep{Bogdanovic2007,Dotti2012,Miller2013}, such as may be the case for J1430+2303. The geometric configuration will become much more complex once the scattering regions (the accretion disks) start to collide (see, e.g, \citealt{Savic2019,Popovic2012}), when the two highly perturbed accretion structures start to align. The interactions between the two SMBHs and their respective disks would lead to many observable signatures that are absent in the case of J1430+2303. Those signatures include, but are not restricted to, flares, a notch in the thermal continuum with a spectral revival at shorter wavelengths (not observed in our VLT spectra but it could very well be out-shined by the host light in the optical band), and a hard X-ray emission with a Wien-like spectrum (T $\sim$ 100~keV) from disk--disk interactions \citep{Roedig2014}. Therefore, we can safely assume that the two potential SMBHs in J1430+2303 are distant enough for our model to be representative, at least at first order.

\begin{figure}
\includegraphics[trim={0 0 0 1.7cm},clip, width=8.8cm]{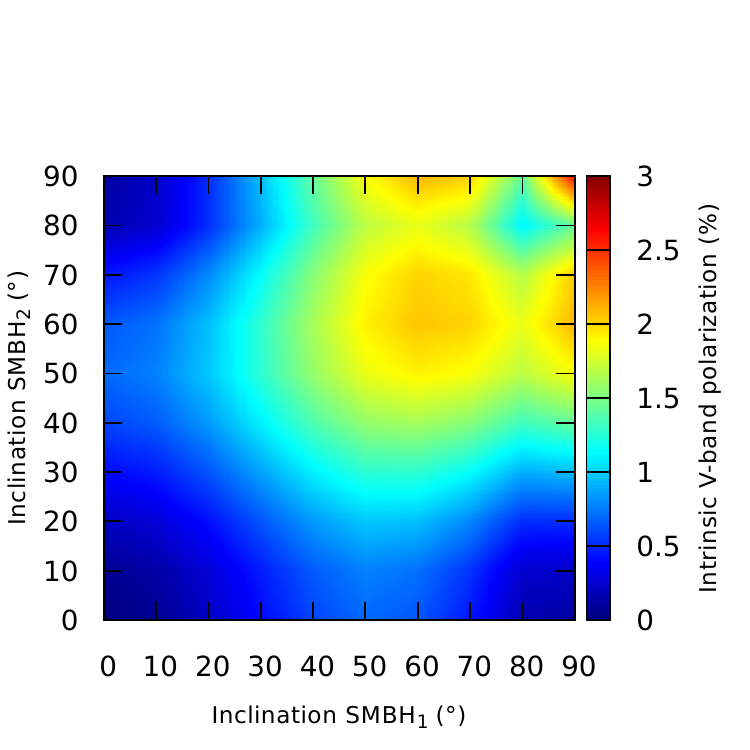}
\caption{Monte Carlo radiative-transfer modeling of the intrinsic V-band polarization 
emerging from a pair of SMBHs with their accretion disks. Each black hole and its
accretion disk has its specific inclination with respect to the observer. The 
total linear polarization is color coded. We note that the BLR, torus, and polar
winds are not included in this simulation.}
\label{Fig:Binary_BH_modeling}
\end{figure}

The V-band polarization resulting from any possible geometrical orientation of two non-interacting SMBHs (and their accretion disk) are plotted in Fig.~\ref{Fig:Binary_BH_modeling}. While thermal polarization is itself unpolarized, scattering inside the accretion disk results in polarization that reaches a few percent as a function of inclination. The maximum polarization degree is found for an orientation of 60$^\circ$, as predicted by \citet{Coleman1991}. However, even in the most favorable parametrizations, the resulting polarization degree never exceeds 2.5\%. 

\begin{figure}
\includegraphics[trim={0 0 0.3cm 0.25cm},clip, width=8.8cm]{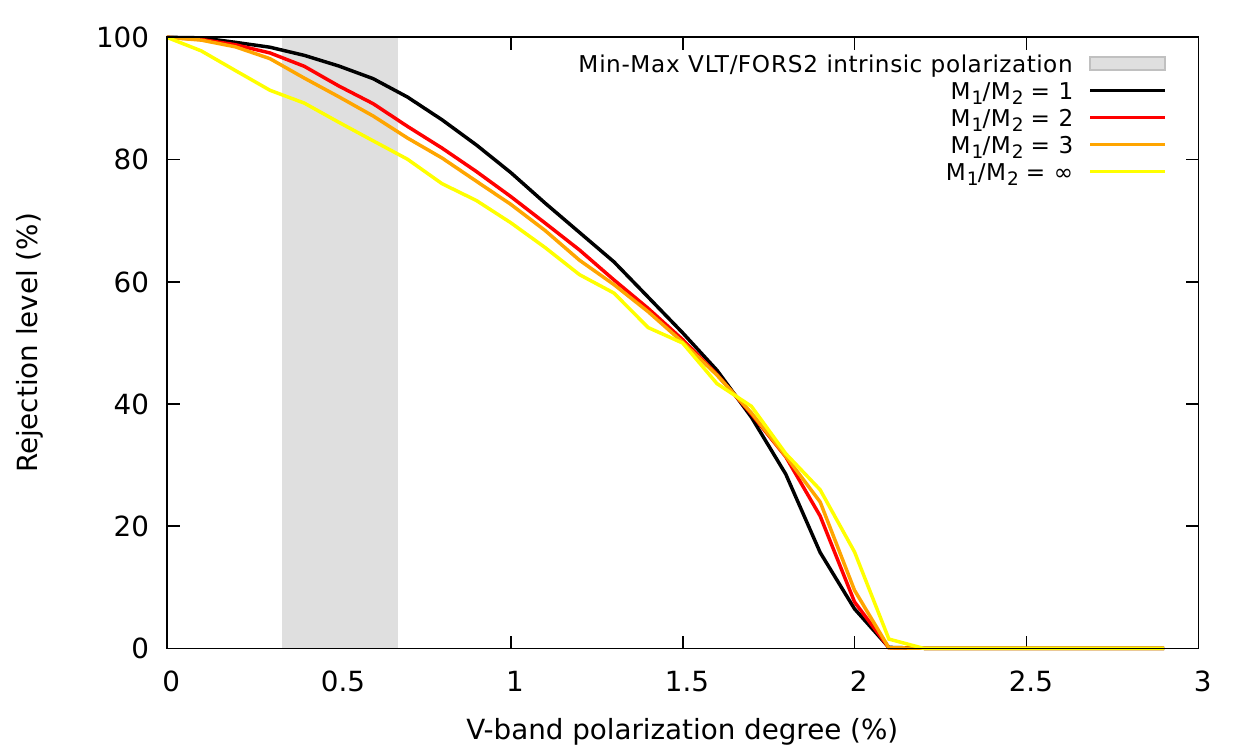}
\caption{Percentage of rejection of the binary SMBH hypothesis as a function of the minimum and maximum intrinsic V-band continuum polarization observed during the period May -- July, 2022. We note that the gray band is different from the one in Fig.~\ref{Fig:Single_BH_modeling} because we no longer plot the three-month averaged V-band value but rather the amplitude of variations in polarization among the three data sets (see text for details).}
\label{Fig:Rejection}
\end{figure}

From the VLT/FORS2 V-band data only, if the system shows variations in $p$, the amplitudes of the variations lie between 0.35\% $\pm$ 0.07\% (July) and 0.49\% $\pm$ 0.07\% (May). Accounting for the 20\% dilution by the host, the potential variation in the (intrinsic) polarization of J1430+2303 ranges from 0.336\% to 0.672\%. From our simulations, the M$_1$ = M$_2$ binary SMBH hypothesis matches polarimetric observations in only 8.8\% of the whole phase space (91.2\% probability of rejection). The very low yet statistically significant polarization degree we measure is therefore rather incompatible with the binary SMBH scenario. Clearly, this conclusion depends on several other parameters. In particular, as suggested by \citet{Jiang2022}, the mass ratio for the two SMBHs might be different. We therefore run {\sc stokes}  again and varied the mass ratio of the two compact objects, with a null-hypothesis that the AGN bolometric luminosity is proportional to the mass (L$_{\rm bol}$ $\propto$ M$^{1.1 \pm 0.3}$, \citealt{Koratkar1991}). The results are presented in Fig.~\ref{Fig:Rejection}. The rejection levels decrease from 91.2\% (M$_1$ = M$_2$) to 84.5\% in the case of M$_1$ = 3 $\times$ M$_2$. From measurements of the linear polarization degree of the continuum only, the probability of having a binary SMBH is about 15\%. Taking into account the lack of variability in photometry and in polarization angle over more than three months, the probability that a binary system lies at the heart of J1430+2303 is likely much smaller.

\section{Discussion}
\label{Discussion}
Since the first paper by \citet{Jiang2022}, J1430+2303 has been observed at radio, optical, and X-ray wavelengths \citep{An2022,Dou2022,Dotti2023}, but our paper provides the only polarimetric counterpart of the object. The lack of variability in photometry we observe is consistent with the findings of the dedicated optical follow-up campaign performed by \citet{Dotti2023}, who found no evidence for a continuing oscillation trend. In fact, during their observation, the luminosity of J1430+2303  decreased by about 0.2 magnitudes or more, down to values comparable to those measured during its 2018 state. The trend observed by \citet{Jiang2022} could very well be due to red noise in the AGN light-curve variability \citep{Vaughan2016,Zhu2020}.

The lack of radio outburst as measured by \citet{An2022} using the VLBI tends to rule out the alternative interpretation, which assumes Lense-Thirring-driven precessing jets \citep{Sandrinelli2016} to explain the luminosity oscillations. The jets, if any, have a projected scale of lower than 0.8~pc and a flat radio spectrum.

Variable X-ray luminosity (up to a factor of 7) is, on the other hand, detected on a timescale of a few days, together with changing properties of the warm absorber \citep{Dou2022}. However, if a broad iron K$\alpha$ emission is indeed detected, no velocity shift or profile change has been measured (further longer X-ray observations are needed). In addition, other scenarios, such as variable obscuration by a clumpy wind, could better explain the variations of the properties of the warm absorber than a binary SMBH. 

When adding polarization into the equation, it becomes clear that the support for a binary SMBH inside J1430+2303 becomes weaker. Our Monte Carlo simulations show that a standard Seyfert-1 AGN model matches the polarimetric and spectroscopic observations  very well. A binary SMBH model is hardly reconcilable with the data. Of course, the real nuclear configuration might be more complex than what we assume, and further modeling is necessary, together with new data. However, there is very little published work providing polarization predictions for binary black holes. Our STOKES model cannot be directly compared to that of \citet{Dotti2022} because (1) the number of scattering regions is larger in our model (torus, winds, etc.) and (2) special relativity is included in the model of these latter authors, while it is not in STOKES. Doppler-boosting due to the motion of the secondary black hole has indeed a major effect on the time-evolution of the polarization properties. If both codes cannot be easily compared, their predictions agree: both seem to disprove the presence of a binary SMBH due to the low polarization degree we observe associated with a lack of time-dependent variation in polarization.

\section{Conclusion}
\label{Conclusion}

The lack of variation of the linear polarized continuum of J1430+2303 and its very small degree are compatible with almost all standard type-1 AGNs. In addition to the main argument of a small polarization degree in the continuum, the similarities between the continuum and H$\alpha$ polarization reinforce our conclusion to reject the binary scenario, although it cannot be completely disproved. If the two SMBHs have a very small inclination with respect to the line of sight of the  observer and if one is circularly orbiting the second, then maybe a binary system can be consistent with our polarization data, but the probability for this geometric configuration remains extremely small.

\begin{acknowledgements}
This research has made use of data from the RoboPol programme, a collaboration between Caltech, the University of Crete, IA-FORTH, IUCAA, the MPIfR, and the Nicolaus Copernicus University, which was conducted at Skinakas Observatory in Crete, Greece. The data in this study include observations made with the Nordic Optical Telescope, owned in collaboration by the University of Turku and Aarhus University, and operated jointly by Aarhus University, the University of Turku and the University of Oslo, representing Denmark, Finland and Norway, the University of Iceland and Stockholm University at the Observatorio del Roque de los Muchachos, La Palma, Spain, of the Instituto de Astrofisica de Canarias. The data presented here were obtained in part with ALFOSC, which is provided by the Instituto de Astrof\'{\i}sica de Andaluc\'{\i}a (IAA) under a joint agreement with the University of Copenhagen and NOT. E.\ L.\ was supported by Academy of Finland projects 317636 and 320045. D.B. and N. M., acknowledge support from the European Research Council (ERC) under the European Unions Horizon 2020 research and innovation programme under grant agreement No.~771282. D.H. is research director at the Fonds de la Recherche Scientifique F.R.S. - FNRS (Belgium). D.H. and D.S acknowledge support from the F.R.S. - FNRS (Belgium) under grants PDR~T.0116.21 and No 4.4503.19.
\end{acknowledgements}

\bibliographystyle{aa} 
\bibliography{bibliography} 

\begin{thebibliography}{59}
\expandafter\ifx\csname natexlab\endcsname\relax\def\natexlab#1{#1}\fi

\bibitem[{{An} {et~al.}(2022){An}, {Zhang}, {Wang}, {Shu}, {Yang}, {Jiang},
  {Dou}, {Pan}, {Wang}, \& {Zheng}}]{An2022}
{An}, T., {Zhang}, Y., {Wang}, A., {et~al.} 2022, arXiv e-prints,
  arXiv:2205.03208

\bibitem[{{Andruchow} {et~al.}(2008){Andruchow}, {Cellone}, \&
  {Romero}}]{Andruchow2008}
{Andruchow}, I., {Cellone}, S.~A., \& {Romero}, G.~E. 2008, \mnras, 388, 1766

\bibitem[{{Angel} {et~al.}(1976){Angel}, {Stockman}, {Woolf}, {Beaver}, \&
  {Martin}}]{Angel1976}
{Angel}, J.~R.~P., {Stockman}, H.~S., {Woolf}, N.~J., {Beaver}, E.~A., \&
  {Martin}, P.~G. 1976, \apjl, 206, L5

\bibitem[{{Antonucci}(1993)}]{Antonucci1993}
{Antonucci}, R. 1993, \araa, 31, 473

\bibitem[{{Ba{\~n}ados} {et~al.}(2018){Ba{\~n}ados}, {Venemans},
  {Mazzucchelli}, {Farina}, {Walter}, {Wang}, {Decarli}, {Stern}, {Fan},
  {Davies}, {Hennawi}, {Simcoe}, {Turner}, {Rix}, {Yang}, {Kelson}, {Rudie}, \&
  {Winters}}]{Banados2018}
{Ba{\~n}ados}, E., {Venemans}, B.~P., {Mazzucchelli}, C., {et~al.} 2018, \nat,
  553, 473

\bibitem[{{Bailer-Jones} {et~al.}(2021){Bailer-Jones}, {Rybizki}, {Fouesneau},
  {Demleitner}, \& {Andrae}}]{Jones2021}
{Bailer-Jones}, C.~A.~L., {Rybizki}, J., {Fouesneau}, M., {Demleitner}, M., \&
  {Andrae}, R. 2021, \aj, 161, 147

\bibitem[{{Berdyugin} {et~al.}(2014){Berdyugin}, {Piirola}, \&
  {Teerikorpi}}]{Berdyugin2014}
{Berdyugin}, A., {Piirola}, V., \& {Teerikorpi}, P. 2014, \aap, 561, A24

\bibitem[{{Blinov} {et~al.}(2021){Blinov}, {Kiehlmann}, {Pavlidou},
  {Panopoulou}, {Skalidis}, {Angelakis}, {Casadio}, {Einoder}, {Hovatta},
  {Kokolakis}, {Kougentakis}, {Kus}, {Kylafis}, {Kyritsis}, {Lalakos},
  {Liodakis}, {Maharana}, {Makrydopoulou}, {Mandarakas}, {Maragkakis},
  {Myserlis}, {Papadakis}, {Paterakis}, {Pearson}, {Ramaprakash}, {Readhead},
  {Reig}, {S{\l}owikowska}, {Tassis}, {Xexakis}, {{\.Z}ejmo}, \&
  {Zensus}}]{Blinov2021}
{Blinov}, D., {Kiehlmann}, S., {Pavlidou}, V., {et~al.} 2021, \mnras, 501, 3715

\bibitem[{{Bogdanovi{\'c}} {et~al.}(2007){Bogdanovi{\'c}}, {Reynolds}, \&
  {Miller}}]{Bogdanovic2007}
{Bogdanovi{\'c}}, T., {Reynolds}, C.~S., \& {Miller}, M.~C. 2007, \apjl, 661,
  L147

\bibitem[{{Coleman} \& {Shields}(1991)}]{Coleman1991}
{Coleman}, H.~H. \& {Shields}, G.~A. 1991, \pasp, 103, 881

\bibitem[{{Dotti} {et~al.}(2022){Dotti}, {Bonetti}, {D'Orazio}, {Haiman}, \&
  {Ho}}]{Dotti2022}
{Dotti}, M., {Bonetti}, M., {D'Orazio}, D.~J., {Haiman}, Z., \& {Ho}, L.~C.
  2022, \mnras, 509, 212

\bibitem[{{Dotti} {et~al.}(2023){Dotti}, {Bonetti}, {Rigamonti}, {Bortolas},
  {Fossati}, {Decarli}, {Covino}, {Lupi}, {Franchini}, {Sesana}, \&
  {Calderone}}]{Dotti2023}
{Dotti}, M., {Bonetti}, M., {Rigamonti}, F., {et~al.} 2023, \mnras, 518, 4172

\bibitem[{{Dotti} {et~al.}(2012){Dotti}, {Sesana}, \& {Decarli}}]{Dotti2012}
{Dotti}, M., {Sesana}, A., \& {Decarli}, R. 2012, Advances in Astronomy, 2012,
  940568

\bibitem[{{Dou} {et~al.}(2022){Dou}, {Jiang}, {Wang}, {Shu}, {Yang}, {Pan},
  {Zhu}, {An}, {Zheng}, \& {Ai}}]{Dou2022}
{Dou}, L., {Jiang}, N., {Wang}, T., {et~al.} 2022, \aap, 665, L3

\bibitem[{{Fossati} {et~al.}(2007){Fossati}, {Bagnulo}, {Mason}, \& {Landi
  Degl'Innocenti}}]{Fossati2007}
{Fossati}, L., {Bagnulo}, S., {Mason}, E., \& {Landi Degl'Innocenti}, E. 2007,
  in Astronomical Society of the Pacific Conference Series, Vol. 364, The
  Future of Photometric, Spectrophotometric and Polarimetric Standardization,
  ed. C.~{Sterken}, 503

\bibitem[{{Gaskell}(2009)}]{Gaskell2009}
{Gaskell}, C.~M. 2009, \nar, 53, 140

\bibitem[{{Goosmann} \& {Gaskell}(2007)}]{Goosmann2007}
{Goosmann}, R.~W. \& {Gaskell}, C.~M. 2007, \aap, 465, 129

\bibitem[{{Heiles}(2000)}]{Heiles2000}
{Heiles}, C. 2000, \aj, 119, 923

\bibitem[{{Hovatta} {et~al.}(2016){Hovatta}, {Lindfors}, {Blinov}, {Pavlidou},
  {Nilsson}, {Kiehlmann}, {Angelakis}, {Fallah Ramazani}, {Liodakis},
  {Myserlis}, {Panopoulou}, \& {Pursimo}}]{Hovatta2016}
{Hovatta}, T., {Lindfors}, E., {Blinov}, D., {et~al.} 2016, \aap, 596, A78

\bibitem[{{IAU Commission 40}(1974)}]{EVPAconv}
{IAU Commission 40}. 1974, in {Transactions of the IAU}, ed. {G. Contopoulos
  and A. Jappel}, {Vol. XVB} ({Dordrecht}: {Reidel}), {p.166}

\bibitem[{{Izzo} {et~al.}(2019){Izzo}, {de Bilbao}, \& {Larsen}}]{Izzo2019}
{Izzo}, C., {de Bilbao}, d.~B., \& {Larsen}, J. 2019, {FORS Pipeline User
  Manual}, VLT-MAN-ESO-19500-4106

\bibitem[{{Jiang} {et~al.}(2022){Jiang}, {Yang}, {Wang}, {Zhu}, {Lyu}, {Dou},
  {Wang}, {Wang}, {Pan}, {Liu}, {Shu}, \& {Zheng}}]{Jiang2022}
{Jiang}, N., {Yang}, H., {Wang}, T., {et~al.} 2022, arXiv e-prints,
  arXiv:2201.11633

\bibitem[{{King} {et~al.}(2014){King}, {Blinov}, {Ramaprakash}, {Myserlis},
  {Angelakis}, {Balokovi{\'c}}, {Feiler}, {Fuhrmann}, {Hovatta}, {Khodade},
  {Kougentakis}, {Kylafis}, {Kus}, {Modi}, {Paleologou}, {Panopoulou},
  {Papadakis}, {Papamastorakis}, {Paterakis}, {Pavlidou}, {Pazderska},
  {Pazderski}, {Pearson}, {Rajarshi}, {Readhead}, {Reig}, {Steiakaki},
  {Tassis}, \& {Zensus}}]{King2014}
{King}, O.~G., {Blinov}, D., {Ramaprakash}, A.~N., {et~al.} 2014, \mnras, 442,
  1706

\bibitem[{{Koratkar} \& {Gaskell}(1991)}]{Koratkar1991}
{Koratkar}, A.~P. \& {Gaskell}, C.~M. 1991, \apjl, 370, L61

\bibitem[{{Lopez-Rodriguez} {et~al.}(2013){Lopez-Rodriguez}, {Packham},
  {Young}, {Elitzur}, {Levenson}, {Mason}, {Ramos Almeida}, {Alonso-Herrero},
  {Jones}, \& {Perlman}}]{Lopez-Rodriguez2013}
{Lopez-Rodriguez}, E., {Packham}, C., {Young}, S., {et~al.} 2013, \mnras, 431,
  2723

\bibitem[{{Madau} {et~al.}(2014){Madau}, {Haardt}, \& {Dotti}}]{Madau2014}
{Madau}, P., {Haardt}, F., \& {Dotti}, M. 2014, \apjl, 784, L38

\bibitem[{{Marin}(2014)}]{Marin2014}
{Marin}, F. 2014, \mnras, 441, 551

\bibitem[{{Marin}(2018{\natexlab{a}})}]{Marin2018}
{Marin}, F. 2018{\natexlab{a}}, \mnras, 479, 3142

\bibitem[{{Marin}(2018{\natexlab{b}})}]{Marin2018b}
{Marin}, F. 2018{\natexlab{b}}, \aap, 615, A171

\bibitem[{{Marin} {et~al.}(2015){Marin}, {Goosmann}, \& {Gaskell}}]{Marin2015}
{Marin}, F., {Goosmann}, R.~W., \& {Gaskell}, C.~M. 2015, \aap, 577, A66

\bibitem[{{Marin} {et~al.}(2012){Marin}, {Goosmann}, {Gaskell}, {Porquet}, \&
  {Dov{\v{c}}iak}}]{Marin2012}
{Marin}, F., {Goosmann}, R.~W., {Gaskell}, C.~M., {Porquet}, D., \&
  {Dov{\v{c}}iak}, M. 2012, \aap, 548, A121

\bibitem[{{Miller} \& {Krolik}(2013)}]{Miller2013}
{Miller}, M.~C. \& {Krolik}, J.~H. 2013, \apj, 774, 43

\bibitem[{{Mortlock} {et~al.}(2011){Mortlock}, {Warren}, {Venemans}, {Patel},
  {Hewett}, {McMahon}, {Simpson}, {Theuns}, {Gonz{\'a}les-Solares}, {Adamson},
  {Dye}, {Hambly}, {Hirst}, {Irwin}, {Kuiper}, {Lawrence}, \&
  {R{\"o}ttgering}}]{Mortlock2011}
{Mortlock}, D.~J., {Warren}, S.~J., {Venemans}, B.~P., {et~al.} 2011, \nat,
  474, 616

\bibitem[{{Nilsson} {et~al.}(2018){Nilsson}, {Lindfors}, {Takalo}, {Reinthal},
  {Berdyugin}, {Sillanp{\"a}{\"a}}, {Ciprini}, {Halkola}, {Hein{\"a}m{\"a}ki},
  {Hovatta}, {Kadenius}, {Nurmi}, {Ostorero}, {Pasanen}, {Rekola}, {Saarinen},
  {Sainio}, {Tuominen}, {Villforth}, {Vornanen}, \& {Zaprudin}}]{Nilsson2018}
{Nilsson}, K., {Lindfors}, E., {Takalo}, L.~O., {et~al.} 2018, \aap, 620, A185

\bibitem[{{Packham} {et~al.}(2011){Packham}, {Young}, \& {Axon}}]{Packham2011}
{Packham}, C., {Young}, S., \& {Axon}, D.~J. 2011, in Astronomical Society of
  the Pacific Conference Series, Vol. 449, Astronomical Polarimetry 2008:
  Science from Small to Large Telescopes, ed. P.~{Bastien}, N.~{Manset}, D.~P.
  {Clemens}, \& N.~{St-Louis}, 405

\bibitem[{{Pacucci} \& {Loeb}(2020)}]{Pacucci2020}
{Pacucci}, F. \& {Loeb}, A. 2020, \apj, 895, 95

\bibitem[{{Panopoulou} {et~al.}(2015){Panopoulou}, {Tassis}, {Blinov},
  {Pavlidou}, {King}, {Paleologou}, {Ramaprakash}, {Angelakis},
  {Balokovi{\'c}}, {Das}, {Feiler}, {Hovatta}, {Khodade}, {Kiehlmann}, {Kus},
  {Kylafis}, {Liodakis}, {Mahabal}, {Modi}, {Myserlis}, {Papadakis},
  {Papamastorakis}, {Pazderska}, {Pazderski}, {Pearson}, {Rajarshi},
  {Readhead}, {Reig}, \& {Zensus}}]{Panopoulou2015}
{Panopoulou}, G., {Tassis}, K., {Blinov}, D., {et~al.} 2015, \mnras, 452, 715

\bibitem[{{Panopoulou} {et~al.}(2019){Panopoulou}, {Hensley}, {Skalidis},
  {Blinov}, \& {Tassis}}]{Panopoulou2019}
{Panopoulou}, G.~V., {Hensley}, B.~S., {Skalidis}, R., {Blinov}, D., \&
  {Tassis}, K. 2019, \aap, 624, L8

\bibitem[{{Pavlidou} {et~al.}(2014){Pavlidou}, {Angelakis}, {Myserlis},
  {Blinov}, {King}, {Papadakis}, {Tassis}, {Hovatta}, {Pazderska},
  {Paleologou}, {Balokovi{\'c}}, {Feiler}, {Fuhrmann}, {Khodade}, {Kus},
  {Kylafis}, {Modi}, {Panopoulou}, {Papamastorakis}, {Pazderski}, {Pearson},
  {Rajarshi}, {Ramaprakash}, {Readhead}, {Reig}, \& {Zensus}}]{Pavlidou2014}
{Pavlidou}, V., {Angelakis}, E., {Myserlis}, I., {et~al.} 2014, \mnras, 442,
  1693

\bibitem[{{Polletta} {et~al.}(2007){Polletta}, {Tajer}, {Maraschi},
  {Trinchieri}, {Lonsdale}, {Chiappetti}, {Andreon}, {Pierre}, {Le F{\`e}vre},
  {Zamorani}, {Maccagni}, {Garcet}, {Surdej}, {Franceschini}, {Alloin},
  {Shupe}, {Surace}, {Fang}, {Rowan-Robinson}, {Smith}, \&
  {Tresse}}]{Polletta2007}
{Polletta}, M., {Tajer}, M., {Maraschi}, L., {et~al.} 2007, \apj, 663, 81

\bibitem[{{Popovi{\'c}}(2012)}]{Popovic2012}
{Popovi{\'c}}, L.~{\v{C}}. 2012, \nar, 56, 74

\bibitem[{{Ramaprakash} {et~al.}(2019){Ramaprakash}, {Rajarshi}, {Das},
  {Khodade}, {Modi}, {Panopoulou}, {Maharana}, {Blinov}, {Angelakis},
  {Casadio}, {Fuhrmann}, {Hovatta}, {Kiehlmann}, {King}, {Kylafis},
  {Kougentakis}, {Kus}, {Mahabal}, {Marecki}, {Myserlis}, {Paterakis},
  {Paleologou}, {Liodakis}, {Papadakis}, {Papamastorakis}, {Pavlidou},
  {Pazderski}, {Pearson}, {Readhead}, {Reig}, {S{\l}owikowska}, {Tassis}, \&
  {Zensus}}]{Ramaprakash2019}
{Ramaprakash}, A.~N., {Rajarshi}, C.~V., {Das}, H.~K., {et~al.} 2019, \mnras,
  485, 2355

\bibitem[{{Roedig} {et~al.}(2014){Roedig}, {Krolik}, \& {Miller}}]{Roedig2014}
{Roedig}, C., {Krolik}, J.~H., \& {Miller}, M.~C. 2014, \apj, 785, 115

\bibitem[{{Rojas Lobos} {et~al.}(2018){Rojas Lobos}, {Goosmann}, {Marin}, \&
  {Savi{\'c}}}]{Rojas2018}
{Rojas Lobos}, P.~A., {Goosmann}, R.~W., {Marin}, F., \& {Savi{\'c}}, D. 2018,
  \aap, 611, A39

\bibitem[{{Sandrinelli} {et~al.}(2016){Sandrinelli}, {Covino}, {Dotti}, \&
  {Treves}}]{Sandrinelli2016}
{Sandrinelli}, A., {Covino}, S., {Dotti}, M., \& {Treves}, A. 2016, \aj, 151,
  54

\bibitem[{{Savi{\'c}} {et~al.}(2019){Savi{\'c}}, {Marin}, \&
  {Popovi{\'c}}}]{Savic2019}
{Savi{\'c}}, D., {Marin}, F., \& {Popovi{\'c}}, L.~{\v{C}}. 2019, \aap, 623,
  A56

\bibitem[{{Serkowski} {et~al.}(1975){Serkowski}, {Mathewson}, \&
  {Ford}}]{Serkowski1975}
{Serkowski}, K., {Mathewson}, D.~S., \& {Ford}, V.~L. 1975, \apj, 196, 261

\bibitem[{{Shannon} {et~al.}(2015){Shannon}, {Ravi}, {Lentati}, {Lasky},
  {Hobbs}, {Kerr}, {Manchester}, {Coles}, {Levin}, {Bailes}, {Bhat},
  {Burke-Spolaor}, {Dai}, {Keith}, {Os{\l}owski}, {Reardon}, {van Straten},
  {Toomey}, {Wang}, {Wen}, {Wyithe}, \& {Zhu}}]{Shannon2015}
{Shannon}, R.~M., {Ravi}, V., {Lentati}, L.~T., {et~al.} 2015, Science, 349,
  1522

\bibitem[{{Simmons} \& {Stewart}(1985)}]{Simmons1985}
{Simmons}, J.~F.~L. \& {Stewart}, B.~G. 1985, \aap, 142, 100

\bibitem[{{Smith} {et~al.}(2002){Smith}, {Young}, {Robinson}, {Corbett},
  {Giannuzzo}, {Axon}, \& {Hough}}]{Smith2002}
{Smith}, J.~E., {Young}, S., {Robinson}, A., {et~al.} 2002, \mnras, 335, 773

\bibitem[{{Strateva}(2004)}]{Strateva2004}
{Strateva}, I.~V. 2004, PhD thesis, Princeton University, New Jersey

\bibitem[{{van Dokkum}(2001)}]{Dokkum2001}
{van Dokkum}, P.~G. 2001, \pasp, 113, 1420

\bibitem[{{van Dokkum} {et~al.}(2012){van Dokkum}, {Bloom}, \&
  {Tewes}}]{Dokkum2012}
{van Dokkum}, P.~G., {Bloom}, J., \& {Tewes}, M. 2012, {L.A.Cosmic: Laplacian
  Cosmic Ray Identification}, Astrophysics Source Code Library, record
  ascl:1207.005

\bibitem[{{Vaughan} {et~al.}(2016){Vaughan}, {Uttley}, {Markowitz},
  {Huppenkothen}, {Middleton}, {Alston}, {Scargle}, \& {Farr}}]{Vaughan2016}
{Vaughan}, S., {Uttley}, P., {Markowitz}, A.~G., {et~al.} 2016, \mnras, 461,
  3145

\bibitem[{{Volonteri} {et~al.}(2015){Volonteri}, {Silk}, \&
  {Dubus}}]{Volonteri2015}
{Volonteri}, M., {Silk}, J., \& {Dubus}, G. 2015, \apj, 804, 148

\bibitem[{{Wang} {et~al.}(2021){Wang}, {Yang}, {Fan}, {Hennawi}, {Barth},
  {Banados}, {Bian}, {Boutsia}, {Connor}, {Davies}, {Decarli}, {Eilers},
  {Farina}, {Green}, {Jiang}, {Li}, {Mazzucchelli}, {Nanni}, {Schindler},
  {Venemans}, {Walter}, {Wu}, \& {Yue}}]{Wang2021}
{Wang}, F., {Yang}, J., {Fan}, X., {et~al.} 2021, \apjl, 907, L1

\bibitem[{{Weingartner} \& {Draine}(2001)}]{Weingartner2001}
{Weingartner}, J.~C. \& {Draine}, B.~T. 2001, \apj, 548, 296

\bibitem[{{Zamfir} {et~al.}(2010){Zamfir}, {Sulentic}, {Marziani}, \&
  {Dultzin}}]{Zamfir2010}
{Zamfir}, S., {Sulentic}, J.~W., {Marziani}, P., \& {Dultzin}, D. 2010, \mnras,
  403, 1759

\bibitem[{{Zhu} \& {Thrane}(2020)}]{Zhu2020}
{Zhu}, X.-J. \& {Thrane}, E. 2020, \apj, 900, 117

\end{thebibliography}

\end{document}